\documentclass[12pt]{article}
    \usepackage{amsmath,amsxtra,amssymb, amsfonts,verbatim }

\usepackage{graphicx,epstopdf}
\newtheorem{thm}{Theorem}
\newtheorem{cor}[thm]{Corollary}
 
\newtheorem{lemma}[thm]{Lemma}

\def\pf{\noindent{\bf Proof:} }

\def\be{\begin{eqnarray}}
\def\ee{\end{eqnarray}}
\def\bee{\begin{eqnarray*}}
\def\eee{\end{eqnarray*}}
\def\bal{\begin{align}}
\def\enal{\end{align}}
\def\pmx{\begin{pmatrix}}
\def\emx{\end{pmatrix}}
\def\bsq{\begin{subequations}}
\def\esq{\end{subequations}}
\def\mm{ \! - \!}

\def\half{{ \tfrac{1}{2} }}
\newcommand{\proj}[1]{ | #1 \kb  #1|}
\def\wtd{\widetilde}
\def\ot{\otimes}
\def\op{\oplus}
\newcommand{\norm}[1]{ \| #1  \|}
\def\id{{\cal I}}
\def\ovb{\overline}
\def\td{\tfrac{1}{d}}
\def\wh{\widehat}
\def\span{\rm span}
\def\range{\rm range}
\def\cB{{\cal B}}
\def\cH{{\cal H}}

\def\ds{\displaystyle}

\def\tr{{\rm Tr} \, }
\def\trp{{\rm Tr} }
\def\bra{\langle}
\def\ket{\rangle}
\def\kb{ \ket \bra }
\def\nn{\nonumber}

\def\qed{ ~~{\bf QED}}

\title{\bf \Large The structure of degradable quantum channels}

\author{Toby S. Cubitt 
\\{\small Department of Mathematics, University of Bristol, University Walk, Bristol BS8 1TW, UK}
\\{\small toby.cubitt@bristol.ac.uk}
\and Mary Beth Ruskai \\{\small  Department of Mathematics, Tufts University, Medford,
MA 02155, USA} \\ {\small marybeth.ruskai@tufts.edu} \and Graeme Smith 
\\ {\small  IBM T.J. Watson Research Center, Yorktown Heights, NY 10598, USA}  \\ {\small graemesm@us.ibm.com}}

\begin{document}

\maketitle

\begin{abstract}

Degradable quantum channels are among the only channels whose quantum
and private classical capacities are known.  As such, determining the structure of 
these channels is a pressing open question in quantum information theory.
We give a comprehensive review of what is currently known about the 
structure of degradable quantum channels,  including a number of
new results as well as alternate proofs of some known results.  
In the case of qubits, we provide a complete characterization of all 
degradable channels with two dimensional output, give a new 
proof that a qubit channel with two Kraus operators is either degradable
or anti-degradable and present a complete description of anti-degradable
unital qubit channels with a new proof.

For higher output dimensions we explore the relationship
between the output and environment dimensions 
($d_B$ and $d_E$ respectively)
of degradable channels.  
For several broad classes of channels we show that they
can be modeled with a environment 
 that is  ``small" in the sense $d_E \leq d_B$.  
Such channels include all those with qubit or qutrit output,
those that map some pure state to an output
 with full rank,   and all those which can be represented using
 simultaneously diagonal Kraus operators, even in a non-orthogonal
 basis.  
Perhaps surprisingly,  we also present examples of degradable channels with 
``large'' environments, in the sense that the minimal dimension $d_E  > d_B$.
Indeed, one can have $d_E > \tfrac{1}{4} d_B^2$.
   These examples can also be used to give a negative answer to the
   question of whether additivity of the coherent information is helpful for establishing
additivity for the Holevo capacity of a pair of channels.

In the case of channels with diagonal Kraus operators, we describe the
subclass which are complements of  entanglement breaking channels.
We also obtain a number of results for channels in the convex hull of
conjugations with generalized Pauli matrices.   However, a number of
open questions remain about these channels and the more general
case of random unitary channels.

~~\end{abstract}

\tableofcontents

\section{Introduction}

In quantum information theory, a quantum channel is represented by a completely
positive, trace-preserving (CPT) map $\Phi$ on a suitable algebra of operators.
 Devetak and Shor \cite{DS} introduced the concept of a  {\em degradable}  channel 
 by combining the classical notion of a degraded broadcast channel with that of the
 complement of a channel. 
A degraded broadcast channel is a single-sender two-reciever broadcast channel in which 
the one receiver can degrade his/her output to simulate the output of the other.  Such channels are
among the few classical broadcast channels for which the capacity region is known \cite{Cov,Cov2}.
Similarly, Devetak and Shor
showed that degradable channels have additive coherent information, so
that their quantum capacity is given by the coherent information for a single
use of the channel.
Furthermore, Yard, Devetak, and Hayden have shown \cite{YDH} that the coherent information of a degradable 
channel is concave as a function of reference state, so that the 
required optimization can be performed efficiently and the capacity problem for such channels
has been completely resolved.  

Before going further, we make these notions explicit.
In the finite dimensional case any completely positive trace-preserving (CPT) map, $\Phi:M_{d_A} \mapsto M_{d_B}$, 
can be represented using an auxiliary space ${\bf C}_{d_E}$ in the form
 \be   \label{rep.anc}
     \Phi(\rho) = \trp_E \, U \rho U^\dag
 \ee
 where $U$ is a partial isometry satisfying $U^\dag U = I_{d_A}$.
The complementary channel $\Phi^C: M_{d_A} \mapsto M_{d_E}$
can then be defined \cite{DS,Hv,KMNR} by taking the partial trace
over the output space $d_B$ so that
\be   \label{comp}
       \Phi^C(\rho) = \trp_B U \rho  U^\dag.
\ee
Physically, the complementary channel captures the environment's view of the channel, and as such it is not 
surprising that its consideration is useful for understanding quantum channel capacities.  

Devetak and Shor call a channel {\em degradable} if there is
 another CPT map $\Psi$ such that 
 \be   \label{degrad}
     \Psi \circ \Phi = \Phi^C.
 \ee
 It is natural to call a channel {\em anti-degradable} if its complement
 is degradable, i.e.,   there is a CPT map  $\Psi$ such that 
$        \Psi \circ \Phi^C = \Phi$. 
Although the complement is only defined up to a partial isometry, this
does not affect the concept of degradability because this map can be
absorbed into the degrading channel $\Psi$.
 
The coherent information of a channel $\Phi$ with respect to a reference state $\rho$ 
was originally defined in terms of a purification.   Here, we find it more useful
to use an equivalent expression involving the complementary channel,
\be   \label{cohref}
I^{\rm coh}(\Phi,\rho) = S\left( \Phi(\rho)\right) - S\left( \Phi^C(\rho)\right) .
\ee
The coherent information of $\Phi$ is  the maximum of \eqref{cohref} over reference states,
\be
I^{\rm coh}(\Phi) = \max_{\rho} I^{\rm coh}(\Phi,\rho).
\ee
The quantum capacity of a channel is given by
\be \label{Eq:MultiLetter}
Q_C(\Phi) = \lim_{n \rightarrow \infty} \frac{1}{n} I^{\rm coh}(\Phi^{\otimes n}),
\ee
as anticipated by  Lloyd \cite{Lloyd} and others \cite{BNS}.   The proof
was completed by Shor \cite{ShorMSRI}, Devetak \cite{Dev} and others \cite{DW}.
When a channel satisfies the additivity condition, 
\be   \label{add}
 I^{\rm coh}(\Phi^{\otimes n}) = n I^{\rm coh}(\Phi),
\ee
the quantum capacity satisfies the simple ``single-letter'' formula
$Q_C(\Phi) =  I^{\rm coh}(\Phi)$.     It was shown in \cite{DS}
that degradable channels satisfy \eqref{add}.     For completeness,
we give a proof of this in Appendix~\ref{appB}.

Though proving (\ref{Eq:MultiLetter}) was a significant step towards
understanding the quantum
channel capacity, it is not known how to cast the quantum capacity of
a general
channel as a finite optimization problem \cite{DSS97,SS07}.  As a
result, little is known about the
quantum capacity of even very basic channels, such as the
depolarizing channel.  Degradable and
anti-degradable channels \cite{FG,WP} are among the few
for which the quantum capacity is known explicitly.
Degradable channels also play a central role in finding bounds on the
quantum capacity for more general channels.
For example, they were used to find good upper bounds on the capacity of
the depolarizing channel \cite{SSW}, especially in the low noise regime.
Moreover, it was recently shown \cite{S07} that for degradable
channels, the
coherent information is also equal to the private classical capacity,
i.e.,
the capacity for transmitting classical information protected
against an eavesdropper in the sense of \cite{C}.

It is well-known that an  anti-degradable channel must have zero
quantum capacity;
as noted in \cite{FG}, this follows from the no-cloning theorem using
an argument that goes back to \cite{BDS}.   A simple analytic
argument has also
been given by Holevo \cite{Hvp}.
Using very different terminology, anti-degradable channels were
considered implicitly
in several earlier papers \cite{BDEFMS97,Cerf,NG}  in which  conditions
were given for a Pauli channel to be anti-degradable.   We provide an
alternate
formulation and proof of these results.   We also show that every
entanglement-breaking
channel is anti-degradable.    Curiously, although the set of degradable
channels is not convex, the set of anti-degradable channels is
convex, as shown in
Appendix~\ref{appendix:Convex}.

Although most channels are neither degradable nor
anti-degradable, the implications for quantum capacity have generated
some interest in identifying those situations in which the degradability
condition \eqref{degrad} holds.  Earlier work has shown that any channel with simultaneously 
diagonalizable Kraus operators is degradable \cite{DS}, as is the amplitude damping channel \cite{FG}.
It was shown in \cite{WP} that any qubit channel with exactly two Kraus operators 
is either degradable or antidegradable, with specific condtions under which each
(or both) of these hold. 
Conditions for the degradability of bosonic Gaussian 
channels were studied in \cite{GH,HvGauss,WP2}, but will not
be considered here.

Roughly speaking, degradable channels are those for which the complement
 is noisier than the original channel, in the sense that the degrading map 
 adds noise to the original channel to generate the complement.     
Since one would expect noisier channels to be associated with larger
environments, it is natural to
guess that one must have $d_E \leq d_B$.   We show that this holds 
if  any pure input has full rank output, as well as in some specific cases.   These
include channels with output dimension of 2 or 3, as well as channels whose
Kraus operators can be simultaneously diagonalized using a pair of left
and right invertible matrices, as discussed in Section~\ref{sect:conjEB}, following ideas
introduced in \cite{WP}.
Therefore, it may be somewhat surprising that we also find a family of counter-examples
which demonstrate that one can have degradable channels with 
$d_E > d_B$ and that this can happen even 
when $d_A = d_B$.\footnote{Some results along these lines have recently been established independently 
by Myhr and Lutkenhaus in their study of symmetric extendable states \cite{ML08}.  Their techniques offer a promising 
direction for further understanding the structure of degradable channels.}

The rest of the paper is organized as follows.  In Section~\ref{sect:size}, we study the size of the environment, beginning with some notation and elementary observations in Section~\ref{sect:pre}.
 Then in Section~\ref{sect:small}  we prove that under a  condition on output rank
 any degradable channel must satisfy $d_E \leq d_B$.   In Section~\ref{sect:large} we present examples of degradable channels not satisfying 
this condition for which  $d_E > d_B$.   
In Section~\ref{sect:qubit}, we give a complete classification of 
 degradable  channels with qubit outputs.    For unital channels 
 mapping qubits to qubits, we  give necessary and sufficient conditions 
 for anti-degradability equivalent to earlier work of Niu and Griffiths \cite{NG}
 and Cerf \cite{Cerf}.     The details and  an alternate 
proof of the results in \cite{WP} for qubit channels with Choi rank 2
are presented in Appendix~\ref{app:qubit}.
In Section~\ref{sect:d3} we show that degradable channels with qutrit outputs must
have $d_E \leq d_B$, but that other results about qubit maps need not extend to
qutrits.    
   In Section~\ref{sect:conjEB}, we study  degradability criteria based on Kraus diagonal conditions, 
   generalizing the results of \cite{DS} and extending some of the ideas in \cite{WP}.  We pay particular attention to channels whose 
complement is entanglement breaking, and show any such channel is degradable. 
We also show that any channel whose Kraus operators can be simultaneously
diagonalized, even if different non-orthogonal bases are used for the input and
output spaces, has $d_E \leq d_B$ and at least one pure input whose output
has full rank.
In Section~\ref{Pauli:diag},  we consider degradability conditions for
 a special type of random unitary channel in which the unitaries
 are restricted to generalized Pauli matrices.    We show that if such a
 channel is degradable, then the unitaries commute and $d_E \leq d_B$.
 In  Section~\ref{sect:last}, we make a few additional observations.
One concerns degradability in a neighborhood of the identity.
   We also observe that the channels introduced in Section~\ref{sect:large}   
   can be used to show that additivity of coherent information for a pair of
   channels need not  imply additivity of the Holveo capacity for the same pair.

We have also included several appendices.   Appendix~\ref{appA} describes Arvseon's commutant
lifting theorem which can be used to define the complement of a channel in
more general and abstract settings.   Appendix~\ref{appB}  contains a proof that
degradability implies additivity of coherent information, while Appendix~\ref{appendix:Convex} shows that the 
set of antidegradable channels is convex.     Appendix~\ref{app:qubit} 
 contains new proofs of some results about
qubit channels.    Appendix~\ref{app:qbasic} introduces some notation and summarizes
basic facts about qubit channels.   An alternate proof of the results in \cite{WP} 
for qubit channels with Choi rank 2 is given in Appendix~\ref{app:qdeg}.
Notation and some  basic results needed for our formulation and proof of
necessary and conditions for a unital qubit channel to be anti-degradable is
given in Appendix~\ref{app:qanti}.   This is followed by analysis of the special
cases of 3 Kraus operators and depolarizing channels in  Appendices~\ref{app:ant2paul}
and \ref{app:antdep} respectively.    The latter shows explicitly that when
 $d_E > d_A$ the degrading map need not be unique.
 Finally, the general case is considered in Appendix~\ref{app:qantipf}.

\section{Size of environment}   \label{sect:size}

\subsection{Preliminaries}   \label{sect:pre}

We will use the term Choi rank of a channel 
 to mean  the rank of its 
Choi Jamiolkoswski state representative $(\id \ot \Phi)(\proj{\beta})$, where 
$|\beta\rangle = \frac{1}{\sqrt{d_A}}\sum_{i=1}^{d_A}|i\rangle|i\rangle$.
This is the same as the minimal number of Kraus operators, or
the size $d_E$ of the smallest 
pure environment that can generate that noise.   Thus one
  must have  $d_E \leq d_A d_B$.  (Note that the Choi rank is {\em not}  the
same as the usual rank of $\Phi$ considered as linear operator on $M_d$.)

In principle, deciding whether or not a channel is degradable is straightforward.
A necessary condition for degradability is that
\be   \label{kercond}
       \ker \Phi \subseteq \ker \Phi^C.
\ee
Thus, if there is a matrix $A \in \ker \Phi$ which is not in $\ker \Phi^C$, the channel
can not be degradable.   Otherwise, when $d_B \leq d_A$,
it suffices to compute $\Psi = \Phi^C \circ \Phi^{-1}$
on $[\ker \Phi]^\perp$ and test $\Psi$ for complete positivity.   In practice, this may
not be so straightforward because composition is the matrix product when $\Phi$
and $\Phi^C$ are represented in some orthonormal bases for $M_d$ in the standard
way (using the Hilbert-Schmidt inner product $\tr A^\dag B$).    However,   testing
for complete positivity requires reshuffling the result into the form
$\sum_{jk} |e_j \kb e_k| \ot \Psi( |e_j \kb e_k| )$.     Furthermore, when $d_B > d_A$, $\Phi$
does not have a right inverse and the degrading map need not be unique.
 In Appendix~\ref{app:antdep} we show that many
     unital qubit channels which are anti-degradable have
     a family of degrading maps rather than a unique degrador.

For $\Phi: M_{d_A} \mapsto M_{d_B} $, the cases in which one of $d_A,
d_B$ or the
Choi rank $d_E$ equal $1$ are all easily treated as follows:
\begin{itemize}

\item When $d_A = 1$, both $\Phi$ and $\Phi^C$ have unique
outputs which we denote $\rho_B$ and $\rho_E$ respectively.
Moreover, $d_B = d_E$ and $\Phi$ is both degradable and
anti-degradable with degrading map
$\Psi :  \gamma \mapsto ( \tr \gamma) \rho_B$ (or $\rho_E$)
completely noisy.
\item When $d_B = 1$, the only possible CPT map is $\Phi = \tr$
which must have $d_A = d_E$ and Kraus operators $|\phi \kb e_k|$.
Then $\Phi^C = \id$ and $\Phi$ is anti-degradable.

\item When $d_E = 1$, any CPT map must have the form
  $\Phi(\rho) = U \rho U^\dag$ with $U^\dag U = I_{d_A}$, which
  implies that $U$ is a partial isometry and $d_A \leq d_B$.
  Then $\Phi^C(\rho) = \tr \rho$ and $\Phi$ is always degradable
  with degrading map $\Psi = \tr$.
\end{itemize}

Implicit in these examples, is the easily verified
fact that $\id^C = \tr$.    We also observe that the situations
in $d_E = 1$ and  $d_B = 1$ are essentially the only ways in which
every pure input has a pure output.
\begin{thm}    \label{thm:pure}
If $\Phi:M_{d_A}  \mapsto M_{d_B}$ maps every pure state
to pure state, then either

(i)  $d_A \leq d_B$ and $\Phi(\rho) = U \rho U^\dag$ with partial isometry $U$ satifying $U^\dag U = I_
{d_A}$
  is always degradable with Choi rank $d_E = 1$, or

(ii)   $\Phi(\rho) =( \tr \rho)\proj{\phi}$ for all $\rho$ is the
completely noisy
  channel which maps a states to a single fixed pure state and is
  anti-degradable.

  \end{thm}
\pf   Let  $U : {\bf C}_{d_Ad_E} \mapsto {\bf C}_{d_B d_E}$ with
$U^\dag U = I_{d_Ad_E}$  be the partial isometry associated with
the representation \eqref{rep.anc}.   If all outputs are pure, then 
$U | \alpha_k\ot e \ket = | \beta_k \ot \gamma_k \ket $ for any
orthonormal basis $\{ \alpha_k \}$ of $M_{d_A}$.
Since $U$ must map orthogonal vectors to orthogonal vectors,
 $\bra \beta_j, \beta_k \ket  \bra \gamma_j, \gamma_k\ket  =
\delta_{jk}$.  Write 
$j \in J^\perp$ when $\bra \gamma_1, \gamma_j\ket  = 0$.
Then
\be
j   \in J^\perp  ~~ \Rightarrow ~~ \Phi:
  \proj{ \tfrac{1}{\sqrt{2}}(\alpha_1 + \alpha_j )}   \mapsto  \half
\proj{\beta_1} + \half  \proj{\beta_j}
\ee
which is pure if and only if $|\beta_j\ket = |\beta_1\ket$.    Thus,
we have
$|\beta_j\ket = |\beta_1\ket ~~ \forall j \in J^\perp$.    For $j
\notin J^\perp$, we
must have $\bra \beta_1, \beta_j \ket = 0$ and
\be   \label{J}
\lefteqn{ j   \notin J^\perp  ~~ \Rightarrow ~~ \Phi:
  \proj{ \tfrac{1}{\sqrt{2}}(\alpha_1 + \alpha_j )} }   \qquad
\qquad \\  \nn & ~ &
   \mapsto  \half \big( \proj{\beta_1} +
   \bra \gamma_1, \gamma_j\ket    |\beta_1\kb \beta_j | +
      \bra \gamma_j, \gamma_1\ket    |\beta_j\kb \beta_1 |  +  \proj
{\beta_j} \big)
\ee
which gives a pure output if and only if  $ |   \bra \gamma_1,
\gamma_j\ket  | = 1$,
or, in other words, $|\gamma_j \ket = e^{i \theta} |\gamma_1 \ket$.

Now, if $J^\perp$ is empty, then $\Phi$ is of the form  (i).
Otherwise, we
can assume that $2 \in J^\perp$ and repeat the argument in \eqref{J} to
conclude that
\bee
  j   \notin J^\perp  \rightarrow    |   \bra \gamma_2, \gamma_j
\ket  | = 1
  \eee
  which gives a contradiction, since $|\gamma_j \ket$ can not be
proportional
  to two orthogonal vectors.   Hence, $\{ j \notin J^\perp \}$ is
empty and and
  $\Phi$ has the form (ii).   \qquad \qed

\subsection{Channels with Small Environment}  \label{sect:small}

In this section we show that if a degradable channel  maps even one pure state
 to an output with full rank, 
then the channel can always be modeled using
an environment no larger than the output space.     We first
prove a more general lemma from which this result follows
immediately.   Although we restrict attention to finite dimensions
we write  ${\cal H}_A$ for ${\bf C}_{d_A} $ and $\cB(\cH_A)$ for $M_{d_A}$, etc.
to emphasize that we consider mappings involving different spaces, 
even when they happen to have the same dimension.

\begin{lemma}   \label{lemma:toby}
Let $\Phi: \cB(\cH_A)\mapsto \cB(\cH_B)$ be a degradable CPT map, and
for a pure state $| \psi_j \ket $ define $B_j = \range  ~ \Phi(\proj{\psi_j})$
and $E_j = \range ~ \Phi^C (\proj{\psi_j})$.    Then $\dim B_j = \dim E_j$.
Moreover, If the vectors 
$ | \psi_1 \ket,  | \psi_2 \ket, \ldots  | \psi_m \ket $ have the property
$\span  \cup_j B_j = \cH_B $, then $\span  \cup_j E_j = \cH_E$.
\end{lemma}

\pf    We can write the spectral decomposition of  each output  as
\be   \label{rep0}
\Phi(\proj{\psi_j} ) = \sum_{k=1}^{r_j} \mu_{jk}^2 \proj{\phi_k^j} 
\ee
with all $\mu_{jk} > 0$ and $|\phi_k^j \ket \in \cH_B$ orthonormal
for each fixed $j$, i.e.,  $\bra \phi_k^j , \phi_{\ell}^j \ket = \delta_{k \ell}$ .    
By  standard purification arguments,
it follows that if  $U: \cH_A \mapsto \cH_{BE}$ is the partial isometry in the 
 representation \eqref{rep.anc} for $\Phi$, one can also find, for fixed $j$,
orthonormal $|\omega_k^j \ket  \in \cB(\cH_E)$ such that
\be  \label{rep1}
       U |\psi_j\ket = \sum_{k=1}^{r_j}  \mu_{jk} \,
           |\phi_k^j \ot \omega_k^j \ket.
\ee 
Note that this implies $r_j = \dim B_j = \dim E_j $.
Now let $\Psi:\cB(\cH_B)\mapsto M_{d_E}$ be the degrading map with environment
${\cal H}_G$ whose representation \eqref{rep.anc} has the operator
$V: \cH_B \mapsto \cH_{EG}$ so that
$V: |\phi \ket \mapsto |\sigma \ket   \in  \cH_{EG}$.
Define $\gamma_k^j \equiv \trp_G \, V \proj{\phi_k^j } V^\dag$.   Then the degradability 
hypothesis implies that for each $j$
\be  \label{degcond}
  \sum_{k=1}^{r_j}   \mu_{jk}^2 \proj{\omega_k^j}  =   \Phi^C(\proj{\psi_j} ) =
  \Psi \circ \Phi (\proj{\psi_j} ) =     \sum_{k=1}^{r_j}  \mu_{jk}^2 \gamma_k^j   .           
\ee
Now, If $\span  \cup_j E_j \neq \cB(\cH_E)$ then there is a vector  $|\omega^\perp \ket  \in \cB(\cH_E)$
  orthogonal to  \linebreak $\span \{ |\omega_k^j \ket  : j = 1 \ldots m, k = 1 \ldots r_j \} $  
  defined in \eqref{rep1}.   
  But then it follows from \eqref{degcond} that 
\be
    0  & = & \sum_{j = 1}^m \sum_{k=1}^{r_j}   \mu_{jk}^2 |\bra \omega^\perp , \omega_k^j  \ket |^2 = 
   \sum_{j = 1}^m  \sum_{k=1}^{r_j}   \mu_{jk}^2  \tr \gamma_k^j \proj{ \omega^\perp} \nn  \\
&    = &
    \sum_{j = 1}^m  \sum_{k=1}^{r_j}    \mu_{jk}^2  \bra \omega^\perp  \gamma_k^j  \omega^\perp \ket 
\ee
But since $\mu_{jk}^2  > 0$ for all $j, k$ and each $\gamma_k^j$ is positive semi-definite,
this implies that  $\bra \omega^\perp  \gamma_k^j  \omega^\perp \ket = 0$ for all $j, k$.
Therefore,
\be
  0  & = &  \trp_E  \tr  \proj{ \omega^\perp}  \gamma_k^j = 
      \trp_{EG} ( \proj{ \omega^\perp} \ot I_G) V \proj{\phi_k^j } V^\dag  \nn \\
      & = &
        \tr \left [ \left(\proj{ \omega^\perp} \ot I_G\right) V \proj{\phi_k^j } V^\dag \left(\proj{ \omega^\perp} \ot I_G\right) \right]\nn \\
     &   =  & {\norm{ (\proj{ \omega^\perp} \ot I_G) V  |\phi_k^j\ket }}^2
\ee
so that $(\proj{ \omega^\perp} \ot I_G) V  |\phi_k^j\ket = 0$ for all $j,k$.   Since the
 hypothesis   $\span  \cup_j B_j = \cB(\cH_B)$ implies that
 any  $|\phi \ket \in \cB(\cH_B)$ can be written as a superposition of $|\phi_k^j \ket$,
 it follows that  $ \bra \omega^\perp |\Psi(\proj{\phi}) |\omega^\perp \ket = 0$
 for any $|\phi \ket \in \cB(\cH_B)$.  Hence  $\cB(\cH_E)=  \span  \cup_j E_j $. \qed

\begin{thm} \label{thm:graeme}
Let $\Phi: \cB(\cH_A)\mapsto \cB(\cH_B)$ be a CPT map with the property that
it has at least one pure state whose image  $\rho = \Phi(\proj{\psi})$ has full rank, i..e,
rank $\Phi(\rho) = d_B$.    Then if $\Phi$ is degradable, $d_E = d_B$.
\end{thm}
\pf  In this case, the hypothesis of Lemma~\ref{lemma:toby} is satisfied with
$m = 1$ so that  $\cH_E= \range ~\Phi^C(\proj{\psi})$. ~~ \qed

For $d_B>2$, it is not hard to find examples of channels $\Phi:M_{d_A}
\mapsto M_{d_B}$ for which no pure input has an output of rank $d_B$,
so that Theorem~\ref{thm:graeme} does not apply. Simply consider a channel
which is a convex combination of strictly fewer than $d_B$ unitary
conjugations, i.e., $\Phi(\rho) = \sum_{k = 1}^\kappa U_k \rho
U_k^\dag$ with $\kappa < d_B$. While it was shown by Devetak and Shor that any such channel is degradable when $\kappa = 2$,
the question of degradability is unresolved in general when $\kappa > 2$.    However,
partial results are given in Section~\ref{Pauli:diag}.

Another example of a channel which has no outputs with full rank is
the Werner-Holevo channel 
$\mathcal{W}(\rho) = \tfrac{1}{d-1}(I - \rho^T)$, for which every pure input 
has output of rank exactly $d-1 $. For $d = 3$, $\mathcal{W} = \mathcal{W}^C$ so that
this channel is both degradable and anti-degradable, as well as an
extreme point of the set of CPT maps.

For output dimension  $d_B = 2,3$ one always has $d_E \leq d_B$ as observed
in  part (i) of Theorem~\ref{thm:qout} for $d_B = 2$ and 
proved in Section~\ref{sect:d3} for $d_B = 3$.

\subsection{Degradable channels with large environment}  \label{sect:large}

We now give an example which shows that one can have $d_E > d_B$ when
$d_B = 2 d_A$.   Let ${\cal N}: M_d \mapsto M_d$ be a CPT map and
define $\Phi:M_d \mapsto M_2 \ot M_d \simeq M_{2d}$ to be the channel 
\be   \label{bigenv}
    \Phi(\rho) =  \half \proj{0} \ot  \id(\rho) +  \half \proj{1} \ot {\cal N}(\rho) = \rho \op {\cal N}(\rho) 
    \ee
where $\id$ denotes the identity channel . 
Then
\be
   \Phi^C(\rho) = \half  \proj{0} \ot \tr \rho +  \half  \proj{1} \ot  {\cal N}^C(\rho) = 
        \half  \tr \rho \op  \half  {\cal N}^C(\rho).
 \ee
A map $\Psi: M_{2d} \mapsto M_{2d}$ 
can be defined by its action on product states  and extended by linearity.   If
\be
    \Psi(\tau \ot \gamma) =   \bra 0, \tau \, 0 \ket {\cal N}^C(\gamma)  \op 
         \bra 1, \tau \,  1 \ket  \tr \gamma, 
\ee
then it is easy to verify that $\Psi \circ \Phi  =  \Phi^C  $ so that $\Phi$ is degradable.
When  ${\cal N}$ has  $d_F$ Kraus operators $A_k$ so that  
${\cal N}(\rho) = \sum_k A_k \rho A_k^\dag$, then the Kraus operators for
$\Phi$ are $| 0 \ket \ot I$ and $  | 1 \ket \ot A_k$ so that it can be represented
with a $d_F + 1$ dmiensional environment.   In particular, when ${\cal N} $
requires the maximum $d_F = d^2$ operators, $d_B  = 2d < d^2 + 1 = d_G$;
therefore, we have a degradable channel 
$\Phi: M_{d_A} \mapsto M_{d_B}$ whose environment  $G$ has larger dimension
than its output space.

One can generalize the channel \eqref{bigenv} as follows. 
 For any $x \geq \frac{1}{2}$, and any channel ${\cal N}:M_d \rightarrow M_d$, 
we can construct a degradable channel
\begin{equation}   
\Phi(\rho) = x \proj{0}\otimes {\cal I}(\rho) + (1-x)\proj{1}\otimes {\cal N}.
\end{equation}
The complementary channel is then
\begin{equation}
\Phi^C(\rho) = x \proj{0}\otimes \tr(\rho) + (1-x)\proj{1}\otimes {\cal N}^C,
\end{equation}
to which $\Phi$ can be degraded using a channel $\Psi$ whose action on product states is
\be
    \Psi(\tau \ot \gamma) =   \tfrac{1-x}{x} \bra 0, \tau \, 0 \ket {\cal N}^C(\gamma)  \op 
        \big[ \tfrac{2x-1}{x}  \bra 0, \tau \,  0 \ket  \tr \gamma,    +  \bra 1, \tau \,  1 \ket  \tr \gamma \big]. 
\ee
In this case, it may be clearer to note that this implies
\be
   \Psi(\gamma_0 \op \gamma_1) =    \tfrac{1-x}{x} {\cal N}^C(\gamma_0)  \op 
        \big[ \tfrac{2x-1}{x}    \tr \gamma_0,    +  \tr \gamma_1 \big]. 
\ee
and with a slight abuse of notation corresponds to a channel with Kraus operators
\bee
  |0 \ket \bra 1| \bra j |, \qquad  
 \sqrt{ \tfrac{1-x}{x}} |1 \ket \bra 0| \otimes C_i  , \qquad 
\sqrt{   \tfrac{2x -1}{x}}|0\ket \bra 0| \bra j|,
\eee
where $C_i$ are the Kraus operators of ${\cal N}^C$ and $j=0\dots d-1$.

It is natural to ask if one must have $d_E \leq d_B$ when $d_A = d_B$?
The answer is no, as shown be the following example.    Let $d_A = 6$
and $d_B = d_E = 3$.   Let $V: {\bf C}_6 \mapsto {\bf C}_9$ be a partial
 isometry whose range is the symmetric subspace of ${\bf C}_3 \ot {\bf C}_3$
 and define a channel $\Phi_2: M_6 \mapsto M_3$ by
 \be
      \Phi_2(\rho) = \trp_E V \rho V^\dag.
 \ee
 Since $V$ maps onto the symmetric subspace of ${\cal H}_B \ot {\cal H}_E$,
$  \Phi_2^C(\rho) = \tr_B V \rho V_\dag =  \Phi_2(\rho)$, so that this channel
is both degradable and anti-degradable.    Now let $\Phi_1$ denote the 
channel defined in \eqref{bigenv} and let $\Phi = \Phi_1 \ot \Phi_2$.
Then $\Phi$ is degradable and has $d_A = d_B = 6d$ but 
$d_E = 3(d^2+1) > 6d = d_B $.

An alternative generalization of \eqref{bigenv} is obtained by constructing degradable channels from
pairs of channels ${\cal M, N}$ for which there exist
channels ${\cal X,Y}$ such that  
\be
    {\cal X} \circ {\cal N} = {\cal M}^C, \qquad    {\cal Y} \circ {\cal M} = {\cal N}^C,
\ee
by letting 
\be   \label{dualenv}
    \Phi(\rho) =  \half \proj{0} \ot  {\cal M} (\rho) +  \half \proj{1} \ot {\cal N}(\rho) =
       {\cal M}( \rho) \op {\cal N}(\rho) .
    \ee
When the environments of $ {\cal M}$ and $ {\cal N}$ have dimensions $d_E$
and $d_F$ respectively, the environment of $\Phi$ has dimension $d_G = d_E + d_F$.
In the example above, ${\cal M} = \id$ is universally degradable since one can choose 
$Y =  {\cal N}^C$ and its complement $\id^C = \tr $ is a universal degrador because
$\tr {\cal N}(\rho) = \tr \rho$.   It is an open question whether or not other such pairs,
which we call ``co-degradable'' exist.
It is plausible that when one map $  {\cal M}$ has Choi rank $d^2$, the 
other map  $ {\cal N}$ must have Choi rank one.   Thus, one might seek additional
examples in which both  $   {\cal M}, {\cal N}$ have Choi rank $ < d^2$.   It
would be interesting to know the optimal dimensions for pairs of 
  co-degradable channels.

  \section{Channels with qubit outputs}  \label{sect:qubit}
  
  
We now consider channels with qubit outputs.  
Wolf and   Perez-Garcia  \cite{WP} showed that every CPT map
$\Phi: M_2 \mapsto M_2$ with
Choi rank  $\leq 2$ is either degradable or anti-degradable.
We present an alternate proof of their result
which exploits the representation of qubit channels introduced in \cite{KR}
and used in \cite{RSW}.   We also show below that no channel with
  qubit output and  Choi rank  larger than $ 2$ can be degradable.
Therefore,  the degradable qubit channels given in \cite{WP} in fact
exhaust all the possibile degradable qubit channels.  The question remains whether there are degradable channels with qubit outputs, but higher dimensional inputs.  We show that this can happen only for input dimension $3$ and, furthermore, up to unitary conjugations of the input and output, such a channel is unique.

\begin{thm}   \label{thm:qout}
  Let $\Phi:M_{d_A} \mapsto M_2$ be a CPT map with qubit output.  
    If $\Phi$ is  degradable, 
  
  (i)   its Choi rank  $d_E$ is at most two, and 
  
  (ii)  its input dimension $d_A \leq 3$.
 
\noindent  Moreover, when $d_A = 3$,  
   up to unitary conjugations on the input and output, 
      \begin{equation}
   \Phi(\rho) = A_0 \rho A_0^\dagger + A_1\rho A_1^\dagger
   \end{equation}
  with
  \begin{eqnarray}
A_0 = \left( \begin{matrix} 1 & 0 & 0\\ 0 & 2^{-1/2} & 0\end{matrix}\right)  \qquad 
A_1=  \left( \begin{matrix} 0 & 2^{-1/2} & 0\\ 0 & 0 & 1\end{matrix}\right),
\end{eqnarray} 
and this channel is both degradable and anti-degradable. 
\end{thm}

\pf   Part (i) follows from Theorem~\ref{thm:graeme} together with Theorem~\ref{thm:pure}. 
In particular, by Theorem~\ref{thm:graeme} if we are to have $d_E>2$, every pure state must 
be mapped to a rank 1 output.  However, in this case the degradability requirement together
with Theorem~\ref{thm:pure}  gives $d_E =1$.  

To prove(ii),   observe that  part (i) implies that we can write
   \begin{equation}
   \Phi(\rho) = A\rho A^\dagger +B\rho B^\dagger,
   \end{equation}
   with $A^\dagger A + B^\dagger B = I_{d_A}$.  Without loss of generality, we may choose
\be
   A & = \left( \begin{matrix} \sqrt{a_1} & 0 & 0 & \ldots & 0 \\ 
        0 & \sqrt{a_2} & 0 & \ldots & 0 \end{matrix}\right) 
        \ee
   so that
   \be
    B^\dagger B = I_{d_A} - A^\dag A & = 
     \left( \begin{matrix} 1-a_1 & 0 & 0 & \ldots & 0\\ 0 & 1 - a_2 & 0 & \ldots & 0 \\
        0 & 0 & 1 & \ldots & 0 \\
        \vdots &    \vdots &  & \ddots & 0 \\
        0 & 0 &   \ldots & 0 & 1 \end{matrix}\right)
\ee
But since $B$ is a $2 \times d_A $ matrix, $B^\dagger B$ can have rank at most two.  
Thus we have a contradiction unless  $d_A \leq 4$.   When
$d_A = 4$, we must also have $a_1 = a_2 = 1$.   To see that $\Phi$
can not be degradable for $d_A = 4$, use the isomorphism 
${\bf C}_4  \simeq  {\bf C}_2 \ot  {\bf C}_2$ and rewrite all matrices in
block form so that $A = \pmx I & 0 \emx,  ~ B = \pmx 0 & I \emx$ 
and $\rho $ has blocks $P_{jk}$.   Then
\bee
    \Phi(\rho) = \pmx P_{11} & 0 \\ 0 & P_{22} \emx   \quad \hbox{but} \quad
        \Phi^C(\rho) = \pmx \tr P_{11} & \tr P_{12}  \\ \tr P_{21} & \tr P_{22} \emx = \trp_2 \,  \rho .
\eee
This will give a contradiction to \eqref{kercond} for a matrix of the form 
$ \pmx 0 & X \\ X^\dag & 0 \emx $ with $\tr X \neq 0$.    Thus, there are
no degradable channels with $d_A = 2$ and $d_B = 4$.
 
 
When, $ d_A = 3$, either $a_1$ or 
   $a_2$ must equal $1$, in order to ensure that the rank of $B^\dag B$ is no greater than 
2.  Without loss of generality, we can assume that
    $a_1 = 1$ and denote $a_2 = a $.  Then it follows that
\begin{equation}   \label{q32Kraus}
A = \left( \begin{matrix} 1 & 0 & 0\\ 0 & \sqrt{a} & 0\end{matrix}\right)
\qquad  \hbox{and}  \qquad
B = U\left( \begin{matrix} 0 & \sqrt{1-a} & 0\\ 0 & 0 & 1\end{matrix}\right)
\end{equation}
for some unitary $U$.
Now, consider the action of $\Phi^C$ on $\proj{0}$ and $\proj{2}$:
\begin{eqnarray}  
\Phi^C(\proj{0}) &=&  \left( \begin{matrix}\tr A\proj{0}A^\dagger & \tr A\proj{0}B^\dagger \\ \tr B \proj{0}A^\dagger & \tr B\proj{0}B^\dagger \end{matrix}\right)  \nn \\
 & = &   \left( \begin{matrix} 1 & 0 \\ 0 & 0 \end{matrix}\right) = \proj{0}. 
 \ee
 \be
\Phi^C(\proj{2}) &=&  \left( \begin{matrix}\tr A\proj{2}A^\dagger & \tr A\proj{2}B^\dagger \\ \tr B \proj{2}A^\dagger & \tr B\proj{2}B^\dagger \end{matrix}\right) \nn  \\
 & =&   \left( \begin{matrix} 0 & 0 \\ 0 & 1 \end{matrix}\right) = \proj{1},  
 \end{eqnarray}
 and compare it to the action of $\Phi$
 \begin{eqnarray}
\Phi(\proj{0}) &= & \proj{0}\\
\Phi(\proj{2}) & =   &U \proj{1} U^\dagger.  
\end{eqnarray}    
Since $\Phi^C(\proj{0})$ and $\Phi^C(\proj{2})$ are orthogonal, if we hope to degrade $\Phi$ to $\Phi^C$, $\Phi(\proj{0})$ and $\Phi(\proj{2})$ must also be orthogonal, which is only the case if $U = I$.

To complete the proof we need to show that when $\Phi$ is degradable
$a = \half$.   
Observe that when  $U = I$  in \eqref{q32Kraus}   $\Phi$ satisfies
\begin{eqnarray}
\lefteqn{\Phi\left( (1-a) \proj{0} - \proj{1} + a\proj{2}\right) }     \qquad \qquad \nn  \\
&=& (1-a)\Phi(\proj{0}) - \Phi(\proj{1}) + a\Phi(\proj{2}) \\  \nn 
 & = & (1-a)\proj{0} - a \proj{1} - (1-a)\proj{0} + a\proj{1}  
  ~ = ~ 0,
\end{eqnarray}
but
\begin{eqnarray}
\bra 0| \Phi^C\left( (1-a) \proj{0} - \proj{1} + a\proj{2}\right)  |0\ket  = (1-a) - a = (1-2a).
\end{eqnarray}
Thus, \eqref{kercond} holds  
only if $a = \half$.   Finally,  observe that when $a= \half$ it is easy to
check that $\Phi = \Phi^C$ so that the channel is both degradable
and anti-degradable with degrading map $\Psi = \id$.
   \qed
   
   The following theorem is due to Wolf and Perez-Garcia \cite{WP}; we present
an alternate proof in Appendix~\ref{app:qdeg}.
In view of part (i) of Theorem~\ref{thm:qout}, their degradability conditions are
necessary as well as sufficient.
\begin{thm}\label{thm:2to2}
{\em (Wolf and Perez-Garcia)}
Up to unitary conjugations on the input and output, the Choi rank two degradable qubit channels
are exactly those of the form
\begin{equation}\label{PhiPG}
\Phi(\rho) = A_+ \rho A_+^\dagger + A_{-} \rho A_{-}^\dagger,
\end{equation}
where 
  \begin{eqnarray}      \label{qextK}
       A_+ & =&   \cos \half v \,  \cos \half u \,   I + \sin \half v \,   \sin \half u  \,  \sigma_z     
          =    \pmx \cos(\frac{1}{2}[v-u]) & 0 \\ 0 & \cos(\frac{1}{2}[u+v])\ \emx
            \nn \\  & ~ &   \\ \nn
         A_- & =& \sin \half v  \,  \cos \half u  \,  \sigma_x - i \cos \half v \,   \sin \half u  \,  \sigma_y  
            =  \pmx 0 &\sin(\frac{1}{2}[v-u])  \\ \sin(\frac{1}{2}[u+v])&  0 \emx   \qquad,
            \end{eqnarray}   
with $|\sin v| \leq |\cos u|$.   
Moreover, when  $|\sin v| \geq |\cos u|$, a channel of the above form 
is anti-degradable.
\end{thm}

\begin{cor}
The degradable qubit channels $\Phi: M_2 \rightarrow M_2$ are, up to unitary conjugations
on the input and output, exactly those of the form given in Eq.~(\ref{PhiPG}) and Eq.(\ref{qextK})
with $|\sin v| \leq |\cos u|$.
\end{cor}
\pf
From Theorem \ref{thm:graeme} we know that any such $\Phi$ can have at most two Kraus operators, which with the
above theorem implies the result.
\qed

Although degradable qubit maps can not have Choi rank greater than 2,
anti-degradable ones can.   Moreover, the set of anti-degradable qubit 
channels is much larger than the expected set of entanglement breaking ones.
The set of anti-degradable unital qubit maps was essentially
 characterized by Cerf \cite{Cerf} and Niu and Griffiths \cite{NG} using a
rather different language, and without distinguishing the subset of
entanglement breaking channels.      We give an alternate formulations and
proof of their result in Appendix~\ref{app:qantipf}.  
\begin{thm}  \label{thm:CNG}
{\em (Cerf, Niu and Griffiths)}
A unital qubit channel with Kraus operators $a_k \sigma_k$ with $\sigma_0 = I$,Ê
$a_0 \geq a_k \geq 0$ and $\sum_k a_k^2 = 1$ is anti-degradable if and only if
\be   \label{CNG}
   a_i^2 + a_j^2 + a_k^2 + a_i a_j + a_i a_k + a_j a_k \geq  \half
\ee
with $i,j,k$ distinct in $\{1,2,3 \}$.
\end{thm}
It was shown in \cite[Appendix A]{KR} that the Kraus operators for any unital qubit 
channel can be chosen to have the form
$a_k U \sigma_k V^\dag$ with $U,V$ unitary and $\sum_k a_k^2 = 1$, 
Thus, Theorem~\ref{thm:CNG}  gives the general result up to unitary conjugations.
Although \cite{Cerf}  considered  only the combination $(i,j,k) = (1,2,3)$  with the
implicit assumption that  $a_0^2$, the weight given to the identity, was larger
than the weight for any other $a_k^2$, 
conjugating with some $\sigma_n$ gives an obvious extension to
arbitrary unital qubit channels.

The general result is more easily stated in a representation introduced in 
\cite{KR}  in which the action of a unital qubit channel 
\be  \label{qumap}
    \Phi:    \half [I + \sum_j w_k \sigma_k ] \longmapsto  \half [I + \sum_k  
\lambda_k w_k \sigma_k)]
\ee
is given by three multipliers $\lambda_k$.   (See Appendix~\ref{app:qbasic}).
In this framework, Theorem~\ref{thm:CNG} can be restated as follows.
\begin{thm}   \label{thm:CNGlambda}
A unital qubit channel is anti-degradable if and only if it  can be
represented using multipliers $\lambda_k$ satisfying the CP condition
$ (1 \pm \lambda_k)^2  \geq  (\lambda_i  \pm \lambda_j)^2$ and the
condition
\be   \label{CNG:abs}
   \sum_{k = 1}^3  \Big( 1 -| \lambda_k |+ \sqrt{(1 - |\lambda_k|)^2 -
 (|\lambda_i| - |\lambda_j|)^2}        \,  \Big) \geq 2 \ee 
\end{thm}

   We can summarize the degradability classification of  channels with qubit outputs as follows
   with the understanding that the conditions are given up to unitary transformation on the
   input and output.
   \begin{itemize}
   
   \item A channel with $d_B = 2$ is {\em both} degradable and anti-degradable if 
      the input dimension $d_A = 1$ or $d_A = 3$.   When $d_A = 2$, it must
      also have two Kraus operators and satisfy  $\sin u = \cos v$ or, equivalently, $u = v + \frac{\pi}{2}$ in 
      the notation of \eqref{ext}.
      
      \item  A channel with $d_B = 2$ is degradable (but not anti-degradable)
       if $d_E = 1$ or if $d_E = 2$
        and $\sin u < \cos v$ in the notation of \eqref{ext}.
        
        \item A channel with $d_B = 2$ is anti-degradable (but not degradable) if $d_E = 2$ and
          $\sin u > \cos v$ in the notation of \eqref{ext}.  
          
             \item  A unital channel with $d_A = d_B = 2$ is anti-degradable
             if it  satisfies \eqref{CNG}.      The subclass which are also EB
             satisfy  $\sum_k |\lambda_k| \leq 1$; however the set of anti-degradable unital qubit
             channels contains many which are not EB, as described in 
             Appendices~\ref{app:ant2paul} to \ref{app:qantipf} 
                 
   \end{itemize}
   
  In the case of unital qubit channels, these classes also have simple
  descriptions in the multiplier picture. 
     
    \section{Channels with output dimension $d_B = 3$} \label{sect:d3}

In this section we prove an analogue of part (i) of Theroem~\ref{thm:qout} 
for  channels with qutrit output.   To do this, we will use Lemma~\ref{lemma:toby}
to draw conclusions about vectors in the union of the ranges of two pure inputs.
We will also need the following complementary lemma to draw conclusion about
vectors in the intersection of the ranges of two pure inputs.
\begin{lemma}   \label{lemma:graeme}
Let $\Phi: \cB(\cH_A)\mapsto \cB(\cH_B)$ be a degradable CPT map, 
with degrading map  $\Psi:   \cB(\cH_B)\mapsto \cB(\cH_E)$.    
For a pure state $| \psi  \ket  \in  \cH_A$ define $B_{\psi} = \range  ~ \Phi(\proj{\psi})$
and $E_{\psi} = \range ~ \Phi^C (\proj{\psi})$.      Then  
$|\phi \ket \in  B_{\psi}$ implies  $ \range  ~ \Psi(\proj{ \phi}) \subset E_{\psi}$.
\end{lemma}

\pf   As in the proof of Lemma~\ref{lemma:toby},  \eqref{rep0} and \eqref{rep1} 
hold   (with the subscript $j$ omitted, as it is now redundant).    Let
$V: \cH_B \mapsto \cH_{FG}  $ be the partial isometry which implements
the representation \eqref{rep.anc} for $\Psi$ so that  $\Psi(\rho) = \trp_G \, V \rho V^{\dag}$.
For $|\phi_k \ket $ the eigenvectors of $\Phi(\proj{\psi})$  let  $|\sigma_k \ket = V  | \phi_k \ket $.
Then the degradability condition implies
 \begin{eqnarray}
 \sum_{k=1}^r \mu_k^2  \proj{\omega_k} & = & \Psi\circ\Phi(|\psi\ket \bra \psi |)
  = \sum_{k=1}^r \mu_k^2 \trp_G  \, \proj{\sigma_k}.
 \end{eqnarray}
 Now suppose $|\omega^\perp\ket$ is orthogonal to $E_{\psi}$.  Then 
  \begin{eqnarray}
 0 & = & \bra\omega^\perp ,  \sum_{k=1}^r \mu_k^2 |\omega_k\ket \bra\omega_k|  \omega^\perp \ket  
 = \sum_k \mu_k^2 \bra\omega^\perp ,  {\rm Tr}_G \, ( |\sigma_k\ket\bra\sigma_k|) \omega^\perp \ket \nn \\
  & = &  \sum_k \mu_k^2  \, \trp_{FG}  ( \proj{\omega^\perp} \otimes I_{G} )
  \, |\sigma_k\ket\bra\sigma_k|  \nn \\
  & = & \sum_k \sum_n \mu_k^2   |\bra \omega^\perp \ot g_n,  \sigma_k \ket |^2
 \end{eqnarray}
 where we used $I_G = \sum_n \proj{g_n}$.  
Since $\mu_k^2 > 0$,  this implies that 
\be
   0 = \bra \omega^\perp \ot g_n,  \sigma_k \ket = \bra \omega^\perp \ot g_n,  V \phi_k \ket 
\ee
 for all $k, n$.
Now let $|\phi\ket = \sum_k \alpha_k |\phi_k\ket$ be an arbitrary vector in $B_{\psi}$.  Then
\begin{eqnarray}
\bra\omega^\perp \ot g_n, V|\phi\ket &=&
     \sum_k \alpha_k \bra \omega^\perp \ot g_n,  V |\phi_k\ket  = 0   \quad \forall ~ n
     \end{eqnarray}
so that 
\be
0 = {\rm Tr} \bra \omega^\perp |{\rm Tr_G}V|\phi \ket\bra\phi|V^\dagger |\omega^\perp \ket
    =   \tr \bra \omega^\perp \Psi(\proj{\phi}) \omega^\perp \ket.
\ee 
Since $\omega^\perp$ was an arbitrary vector in $E_{\psi}^\perp$, this proves that
$ \Psi(\proj{\phi}) \subseteq  E_{\psi}$.  ~~
\qed

\begin{thm}
  Let $\Phi:M_d \mapsto M_3$ be a CPT map with qutrit output. If  $\Phi$ is
  degradable, then its Choi rank is at most three.
\end{thm}    
 \pf  Let $r_{\max}=\max \{ \hbox{rank} \Phi(\proj{\psi}) :   {|\psi \ket}  \in {\bf C}_{d_A} \} $
  denote the maximum output rank of the channel over all pure-state
  inputs in $  \mathcal{H}_A$.   If $r_{\max}=3$ the result holds by
 Theorem~\ref{thm:graeme}; and if $r_{\max}=1$ the result
follows from Theorem~\ref{thm:pure} as for qubits.
  Thus, we can assume $r_{\max}=2$.     Fix a  $|\psi_1 \ket$
  such that  {$r_1 =  \hbox{rank} ~ \Phi(\proj{\psi_1})  = 2$}.     As in 
   Lemma~\ref{lemma:graeme},  let  $B_{\psi} = \range  ~ \Phi(\proj{\psi })$
and $E_{\psi} =   \range ~ \Phi^C (\proj{\psi})$.
 If  $B_{\psi} \subseteq  B_1 $  for all  $|\psi \ket \in \cH_A$, then we have a 
  qubit output embedded
  in a qutrit  space and the result follows from Theorem~\ref{thm:qout}.
 Otherwise there is a second vector $|\psi_2 \ket$ 
 for which $B_2 \nsubseteq  B_1 $.   
 If $r_2 = 1$, one can find a superposition  
 $|\psi \ket = a |\psi_1\ket + b |\psi_2 \ket$  whose output has rank 2
 and for which  $B_{\psi} \nsubseteq  B_1$.\footnote{To see this
 write
   $|\psi_1 \ket = \mu_1 |\phi_1 \ot f_1 \ket +  \mu_2 |\phi_2 \ot f_2 \ket $
 with $\phi_j$ and $f_j$ respectively orthogonal for $j = 1,2$.
If  $|\psi_2 \ket =  |\phi_3 \ot f_3 \ket $, then $a |\psi_1 \ket  + b |\psi_2 \ket $
  must have rank  $\leq 2$, because rank 3 is excluded by assumption.
  Roughly, the only superposition which could yield a state of rank 1 must
  have the form    $a |\psi_1 \ket  - b  |\phi_j \ot f_j \ket $; however, the 
  assumption  $B_2 \nsubseteq  B_1 $ precludes $| \phi_3 \ket = | \phi_j \ket $ for $j = 1,2$.
  For a precise argument, write
  $  |\psi_2 \ket =  t_1 |\phi_1 \ot f_3 \ket +    t_2 |\phi_2 \ot f_3 \ket  + t_3 |\wh{\phi}_3 \ot f_3 \ket $
  with $\bra \wh{\phi}_3 , \phi_j \ket = 0 ~~ j = 1,2$.   Let
  \bee
   |\psi\ket \equiv  a |\psi_1 \ket  + b |\psi_2 \ket  =  |\phi_1 \ot g_1 \ket  + |\phi_2 \ot g_2 \ket  +  |\wh{\phi}_3 \ot g_3 \ket 
  \eee
  with unnormalized vectors $g_j = a \mu_j f_j + b t_j f_3$ for $j =1,2$ and $g_3 = b t_3 f_3$.   
  Then the density matrix $\trp_E \proj{\psi}$ can be represented by the $3 \times 3$ matrix with 
  elements $\bra g_j , g_k \ket$.   If this has rank 1, then the subdeterminants  
  $\bra g_j , g_j \ket  \bra g_k , g_k \ket  - | \bra g_j , g_k \ket |^2 = 0$.   But this implies
  $g_3 = c g_j$ for $j = 1,2$ which implies $f_3 = c^\prime f_1$ and $   f_3 = c^{\prime \prime} f_2$
  which is impossible since $\bra f_1, f_2 \ket = 0$.}
  
     Thus we have reduced the problem to the case in which $\dim B_2 = \dim B_1 = 2$,
     and $B_1 \neq B_2$.    The assumption that $d_B = 3$ then implies that
     $\span B_1 \cup B_2 = {\cal H}_B$.   Moreover, $\dim  {\cal H}_B = 3$ implies
     that $B_1 \cup B_2 \neq \emptyset$.     It follows from Lemma~\ref{lemma:toby} that
    $\dim E_1 = \dim E_2 = 2$ and    $\span ~ E_1 \cup E_2 = {\cal H}_E$.
    Now let $| \phi \ket \in B_1 \cap B_2$.   By Lemma~\ref{lemma:graeme} 
    $\Psi(\proj{\phi}) \in E_1$ and $\Psi(\proj{\phi}) \in E_2$.  Therefore,
    $E_1 \cap E_2$ is non-empty, and
    \be
        \dim  {\cal H}_E = \dim E_1 + \dim E_2 - \dim E_1 \cap E_2 \leq 2 + 2 -1 = 3.  \qquad \qed
    \ee
        

Unlike the case of qubits, not every map $\Phi: M_3 \mapsto M_3$  with
Choi rank 3 is either degradable or anti-degradable.   A specific class
of examples is given in Corollary~\ref{cor:d3}.
For $\Phi: M_4 \mapsto M_4$ one can obtain a simpler example.
Let $\Phi_1$ be a  qubit channel which is degradable (but not 
    anti-degradable) and $\Phi_2$ be a  qubit channel which is 
    anti-degradable (but not  degradable).   Then  $\Phi = \Phi_1 \oplus \Phi_2$
    has 4 Kraus operators, but is neither degradable nor anti-degradable.

 \section{Kraus diagonal conditions} \label{sect:conjEB}
   
  Devetak and Shor  \cite{DS}, showed that any channel with simultaneously
   diagonalizable Kraus operators is degradable.   These are often
   called ``diagonal channels'' following terminology introduced in
   \cite{LS} and followed in \cite{King6}.   However,  we prefer the term 
   ``Hadamard'' used in \cite{KMNR} or ``Kraus diagonal'' to avoid 
   confusion with channels represented by a diagonal matrix when thought of as a linear operator on the
vector space of density operators.  King \cite{King6} showed that
 a CP map  has diagonal Kraus operators if and only if it can be represented in
 the form   $\rho \mapsto  H * \rho$ with $H$ positive semi-definite
  where $*$ denotes Hadamard (or pointwise) multiplication.    It
 is easy to invert  $H*\rho$ since  $J* H* \rho = \rho$ when $J$ has
 elements   $1/h_{jk}$.
 
 A channel is equivalent to one with diagonal Kraus operators 
 if there are unitary
 $U,V$ such that  $A_m = U^\dag D_m V$ where  $D_k$ is diagonal with
 elements $a_{jm}$ on the diagonal.     Thus, in essence, the operators
 $A_m$ have a simultaneous SVD in which one has dropped the usual 
 requirement of positive elements on the diagonal.   
  The matrix $H$ then has elements
 $h_{jk} = \sum_m a_{jm} \ovb{a}_{km} $.     Thus
 \be
      \Phi(\rho)  = \Gamma_{U^\dag} \big(H* \Gamma_V(\rho)\big)
 \ee
 where  $\Gamma_V(\rho) = V \rho V^\dag$.
   
    In \cite{WP}, Wolf and Perez-Garcia  introduced the notion of ``twisted diagonal''  
  for $\Phi:M_{d_A} \mapsto M_{d_B} $ with Kraus operators $A_m$.
  They considered only $d_A$ = $d_B$ and required that there exist
   invertible  $Y, X$  such that    $Y A_m  X$ is diagonal.   It is not 
   hard to see that this can be extended to channels with $d_A \le d_B$
   for which  $Y$  and $X$ have left and right inverses satisfying  
   $Y_L^{-1} Y = I_A$ and $X  X_R^{-1} = I_A$ respectively.
     The main idea is that $\Phi$ can then be written as a composition 
 using single conjugations and Hadamard multiplication, i.e,
$ \Phi(\rho)  = \Gamma_{Y} \big(H* \Gamma_X(\rho)\big) $ where
$\Gamma_Y(A) = Y A Y^{\dag}$.   Since these maps are  
 easy to invert, Wolf and Perez-Garcia could then give a simple test for  
 degradability
 of twisted diagonal channels.    They also showed that  a  channel 
  $\Phi:M_{d} \mapsto M_{d}$ with Choi rank two    
 is twisted diagonal if one of the
 Kraus operators has rank $d_A$.  
 The extreme amplitude-damping channel with Kraus operators  
  $|0\kb 1|$ and  $|0\kb 0|$ is not twisted diagonal because a matrix of
  the form  
  $A = \big( \begin{smallmatrix} 0 & a \\  0 & 0 \end{smallmatrix} \big)$ can not be further reduced.
  
   A large class of degradable channels that are twisted diagonal 
   can be constructed by considering the complements of  
   entanglement breaking  (EB)  maps.    It is convenient to begin with
   the map $\Phi^C: M_{d_A} \mapsto M_{d_E} $ and recall that \cite{HSR}
  a CP map  $\Phi^C$ is EB if and only if its Kraus
   operators can be chosen to have rank one, so that
   \be
       \Phi^C(\rho) = \sum_k A_k \rho A_k^\dag = \sum_k  \proj{x_k} \bra w_k, \rho  \, w_k \ket
   \ee
     with $A_k = |x_k \kb w_k |$.  It was shown in \cite{Hv,KMNR}   that
     the complement  $\Phi : M_{d_A} \mapsto M_{d_B} $  has the form
     \be   \label{psdiag}
       \Phi(\rho) =  \sum_m F_m \rho F_m^\dag =
       \sum_{jk} |e_j \kb e_k|   x_{jk}   \bra w_j, \rho  \, w_k \ket
     \ee             
  with $x_{jk} =  \bra x_j , x_k \ket$.   Moreover,
   the Kraus operators of $\Phi$ have the pseudo-diagonal form
   $F_m =  \sum_k c_{km} |e_k \kb w_k |$, where $C = (c_{km})$ satisfies $(CC^\dag)_{jk} = \bra x_j , x_k \ket$.
      We call this pseudo-diagonal
   because the vectors $|w_k \ket$ need not be orthonormal, although the
   $| e_k \ket$ are orthogonal.  Note that if   $W$ is the matrix with
   elements  $ \bra w_j, \rho  \, w_k \ket$, then $\Phi(\rho)$ is represented
   by the Hadamard product $X*W$.  It was also shown in \cite{Hv,KMNR} that
   a channel has the form \eqref{psdiag} if and only if it is the complement of
   an EB map.   A pseudo-diagonal channel is a special case of a twisted
   diagonal channel with $Y$ unitary.    It follows from Theorem 6 in \cite{HSR}
   that  $d_E \leq d_B$.  (In our notation $d_B$ is  the dimension of
   the environment of the EB channel  $\Phi^C$.
   Actually, this result is stated only for $d_A = d_E$ but
   easily generalizes to $d_B \geq \max\{ d_A, d_E \}$.)
   
    \begin{thm}   \label{thm:pseudo}
   Every pseudo-diagonal channel is degradable. Equivalently, every entanglement breaking 
channel is anti-degradable.  
       \end{thm}      
\pf   Let  $\Psi$ be the CP map with Kraus operators  
    $G_k = \frac{1}{\| x_k \|}  |x_k \kb e_k |$.  
    Then it follows immediately from \eqref{psdiag}
    that 
    \be
        \Psi \circ \Phi(\rho) & = & \sum_{\ell}    
            \sum_{jk}  \delta_{j \ell} \delta_{k \ell }  \,  \proj{x_{\ell}}  \,
                 \frac{  \bra x_j , x_k \ket }{ \bra x_k , x_k \ket  }  \, 
               \bra w_j, \rho  \, w_k \ket  \\
               & = &   \sum_{\ell}     \,  \proj{x_{\ell}}  \,    \bra w_{\ell}, \rho  \, w_v \ket  = \Phi^C(\rho).  \qed
    \ee

      
\begin{thm}
If $\Phi: M_{d_A} \mapsto M_{d_B}$ is twisted diagonal with $d_A = d_B$,
then $d_B \ge d_E$ and there is a pure 
state such that the rank of the output $\Phi(\proj{\psi})$ is $d_B$.
\end{thm}
\pf    The $d_E$ Kraus operators  $A_m$ in a minimal set are linearly
independent because they are eigenvectors of the CJ matrix.   
For  $d_A = d_B$ left and right inverses exist if and only if $X,Y$ are
invertible.    Thus  $A_m = Y D_m X$ with $X,Y$ invertible and 
$D_k$ diagonal with  $a_{jm}$ on the diagonal.   
The vectors  ${\bf a}_m$ are also
linearly independent, which implies that   $d_E \leq d_B$.
Let ${\bf a}_m$ denote the vectors with elements $a_{jm}$ and
$H = \sum_m {\bf a}_m   {\bf a}_m^\dag $, and note that it has
  rank $d_E$.  Then
\be
     \Phi(\proj{\psi}) = Y \big[ H* (X \proj{\psi}X^\dag) \big]Y^\dag .
\ee
 Since $X$ is
invertible, one can find   $|\psi_1 \ket $ such that  $X |\psi_1 \ket $  is
proportional to  $( 1, \dots, 1)^T$.  
Then  $H* (X \proj{\psi}X^\dag)  =  c H$ for some constant $c$.
Since $Y$ is invertible, it does not affect the rank, so  $\Phi(\proj{\psi}) $
has rank $d_E$.   \qed

It is curious that we could not show directly that there is an
input whose output has full rank, and apply Theorem~\ref{thm:graeme}.
Instead, we first showed that  $d_E \leq d_B$ and used this to conclude
that a pure state with full rank output exists.
 In the case of pseudo-diagonal channels, we have also been unable to
 show that there is a pure input whose output has full rank.
    It would be enough to show that one can find
a $\psi$ such that  $\bra w_k , \psi \ket \neq 0$ for all $k$.

\section{Random unitary and Pauli diagonal channels}  \label{Pauli:diag}

We now explore the conditions for the degradability of random unitary channels.
A random unitary channel $\Phi: M_d \mapsto M_d$ is a convex 
combination of unitary conjugations, i.e.,
 \be   \label{randu}
     \Phi(\rho) =   \sum_{k=1}^{\kappa}  a_k U_k \rho U_k^\dag
 \ee
 with  each  $a_k \geq 0$ and $\sum_k a_k = 1$.
When there are precisely $\kappa$ distinct unitaries, a pure input can have
output of rank at most $\kappa$.    If there are  $d$ or more unitaries, one
would expect that one can always find at least one pure input whose output
has rank $d$.   If so, one can apply Theorem~\ref{thm:graeme}.
However, we have not found a proof of this, and one can    
easily construct examples for which some inputs have lower rank.   
Nevertheless, one can show directly that for an important
subclass of random unitary channels, degradability implies 
$d_E \leq d_B = d_A$.

Let $X$ and $Z$ denote the matrices whose action on the standard basis
is $X |e_k\ket = |e_{k+1} \ket $ and $Z |e_k\ket =  e^{i2 \pi k/d} |e_k \ket $.
The  unitary matrices $X^j Z^k$ are called generalized Pauli matrices
and give a projective representation of the Weyl-Heisenberg group.  
Let  $V_m$ denote some ordering of  $X^j Z^k$ with $V_0 = I$.
Then $\tr V_m^\dag V_m = d \delta_{mn}$ and 
 one can write any density matrix in $M_d$ as
\be  \label{dbloch}
    \rho = \td\big[I + \sum_{k=1}^{d^2-1} v_k V_k \big]
\ee
 with $v_m = \tr V_m^\dag \rho$.    One can show that
that   $|v_m| \leq 1$ and $ \sum_m |v_m|^2 = d-1$.   Moreover, when
 $\rho$ is pure 
  $|v_m| = 1$ for exactly $d-1$ of the $v_m$ and the rest are zero.
  For details see \cite{FH,KMNR,NR}.

 We now restrict attention to channels $\Phi$ of the form \eqref{randu} where each 
 unitary is one
of the generalized Pauli matrices.      Any such channel is equivalent via unitary
conjugation to a channel with $a_0 \neq 0$, and we will assume that this holds.
   In general, if the $V_m$ corresponding to the remaining 
non-zero $a_m$ do not commute, we do not expect the channel to be
degradable.      Theorem~\ref{thm:Pauli}, together with Corollary~\ref{cor:PauliNonCom} makes this intuition precise.
 The channel $\Phi$ is represented by
the  matrix with elements $\tr V_n^\dag \Phi(V_m) = \phi_m \delta_{mn}$ 
where  $\phi_m = \sum_n \xi_{mn} a_n$ and  
 $\xi_{mn} = \td \tr V_m V_n V_m^\dag V_n^\dag $ is a $d^\mathrm{th}$ root 
 of unity arising from the Weyl-Heisenberg 
 commutation relations.   Therefore,  
  $|\phi_m| \leq 1$ with equality if and only if  $\xi_{mn} = +1$ 
whenever $a_n \neq 0$.  
 Since $\Phi$ is represented by
a diagonal matrix, we call such channels {\em Pauli diagonal}.    The
 effect of $\Phi$  on a density matrix represented in the form \eqref{dbloch} is simply  to map 
 $v_m \mapsto \phi_m v_m$.   
 
 \begin{thm} \label{thm:Pauli}
Let $\Phi$ be a channel of the form \eqref{randu} with each $U_k$ one of
the generalized Pauli matrices.
 If  $a_0 \neq 0$ and for some $m > 0$, $a_m \neq 0$ and $0 < |\phi_m| < 1$,
 then $\Phi$ is not degradable.
 \end{thm} 
 \pf  For simplicity, we first consider the case when $V_m$ has order $d$, i.e.,  
 $V_m^d  = I$ but
  $V_m^{\kappa} \neq I$ for any positive integer $\kappa < d$.
 Then  
 $\ds{ \rho = \td \sum_{k = 0}^{d-1} V_m^k }$ projects onto an eigenstate
 of $V_m$ and is, hence, positive semi-definite.   
 We will show that
 $\Phi$ is not degradable by showing that  $ \Phi^C \circ \Phi^{-1}(\rho)$
 is not positive semi-definite.   (If some $\phi_n = 0$, then $\Phi$ is not
 degradable unless  $ \Phi^C(V_n) = 0$ also.   When this happens,  
 it suffices to invert $ \Phi$ on $\ker(\Phi)^\perp$.)
 
 Using an obvious abuse of notation, we find
 \bee
      \Phi^{-1}(\rho) = \td \sum_{k = 0}^{d-1}   \phi_{V_m^k}^{-1} V_m^k .
 \eee
 Now, it suffices to consider the following $2 \times 2$ submatrix
 of  $(\Phi^C \circ \Phi^{-1})(\rho)$,
 \be
   \pmx   a_0    & \sqrt{a_0 a_m}  \tr \Phi^{-1}(\rho) V_m^\dag \\
    \sqrt{a_0 a_m}  \tr  V_m \Phi^{-1}(\rho)   & a_m \tr V_m \Phi^{-1}(\rho) V_m^\dag \emx
    = \pmx  a_0  &  \sqrt{a_0 a_m} \phi_m^{-1} \\
         \sqrt{a_0 a_m}  \ovb{\phi}_m^{-1}   &   a_m \emx,
 \ee
 the determinant of which is $a_0 a_m (1 - |\phi_m|^{-2}) < 0$.  
 
 If $d$ is not prime, e.g., $d = d_1 d_2$ and $V_m^{d_2} = I$, then 
 $V_m =  V_{m^*}^{d_1}$  for some $m^*$.   In that case, we can 
apply the same argument to   $\ds{ \rho = \td \sum_{k = 0}^d V_{m*}^k }$. ~~~ 
 \qed
 
 \begin{cor}\label{cor:PauliNonCom}
Let $\Phi$ be a Pauli-diagonal channel with $a_l \neq 0$ and $a_k \neq 0$ and $V_lV_k \neq V_kV_l$.  Then $\Phi$ is not 
degradable. 
 \end{cor}
\pf
Assume, without loss of generality, that $V_0 =I$ and $a_0 \neq 0$.
First note that if there is some $a_m\neq 0$ such that $\phi_m=0$, the channel cannot be degradable,
since
\begin{equation}
\Phi(V_m) = 0,
\end{equation}
but
\begin{equation}
\bra{0}|\Phi^C(V_m)|m\ket = \sqrt{a_m} \tr V_m V_m^\dag = d\sqrt{a_m}\neq 0.
\end{equation}
But the usual observation that ${\rm Ker} \Phi^C \subset {\rm Ker} \Phi$ is required for degradability shows that the channel could not be degradable.

If we can also rule out the possibility that $|\phi_m| =1$ for all $m$ with $a_m \neq 0$, we will be able to 
use Theorem \ref{thm:Pauli} to establish the result.  But, recall that $|\phi_m| = 1$ only if $\xi_{mn} = 1$
for all $a_n \neq 0$, so that in this case all the $V_n$ with nonzero $a_n$ must commute. 
\qed

 \begin{cor}
 Let $\Phi$ be a channel of the form \eqref{randu} with each $U_k$ one of
the generalized Pauli matrices and Choi rank $> d$, i.e., $\left|\{a_m \neq 0\}\right| >d$.
Then $\Phi$ is not degradable.
 \end{cor}
\pf
 Since the generalized Paulis are linearly independent, any mutually commuting subset can contain at most $d$ elements, so that there must be at least two $V_n$ with nonzero $a_n$ that don't commute, which by the previous corollary establishes the result.
\qed

 The following corollary is of interest because is shows that for $d > 2$,
  there are channels with exactly $d$ Kraus operators which are neither
  degradable nor anti-degradable.       
 \begin{cor}   \label{cor:d3}
Let   $\Phi: M_3 \mapsto M_3 $ be the channel
 $\Phi(\rho) =  a_0 \rho +  a_1 X \rho X^\dag + a_2  Z \rho Z^\dag$
with $a_0, a_1, a_2$ strictly positive and at least two unequal.
Then $\Phi$ is neither degradable nor anti-degradable.
 \end{cor}
 \pf  Since $X$ and $Z$ do not commute, it follows from Corollary~\ref{cor:PauliNonCom} that $\Phi$ is not degradable.
 Indeed,  this holds even when $a_0 = a_1 = a_2 = \tfrac{1}{3}$.   To show 
 that $\Phi$ is not anti-degradable, we show that $\Phi$ has strictly positive
 coherent information  $I_Q(\Phi)$ by considering its action on a maximally 
 entangled state $|\beta \ket$.   One finds that
 $\Phi(\proj{\beta}) = \Phi(   \tfrac{1}{3} I) =  \tfrac{1}{3} I$ and that
 $(\Phi \ot \id)(\proj{\beta})$ has eigenvalues $a_0, a_1, a_2$.  To see the
 latter it suffices to observe that the states  
 $\{ |\beta \ket, ~  X  |\beta \ket , ~Z |\beta \ket \}$ are mutually orthogonal.
 But this holds since, e.g.,
 \bee
     \bra \beta, (X \ot I) \beta \ket = \trp_{12}  (X \ot I) \proj{\beta} =
          \tr X  ( \tfrac{1}{3} I) = 0
 \eee
Thus, we find  $I_Q(\Phi) \geq \log 3 + \sum_k a_k \log a_k > 0$ unless
$a_0 = a_1 = a_2 =  \tfrac{1}{3}$.       \qed

  We have not resolved the question of whether or not the channel with
  all $a_k = \tfrac{1}{3}$ is anti-degradable.   A more interesting question
  is whether or not the degradability result (which holds even when all
  $a_k = \tfrac{1}{3}$)  remains true when $X$ is replaced by an arbitrary
  unitary operator which does not commute with $Z$.
  
 \section{Additional remarks}    \label{sect:last}
  
 It was also shown in \cite{WP} that
{\em any} channel  $\Phi$ with Choi rank two
that is sufficiently close to the identity map $\id $ is degradable. 
It is worth remarking  this is not the same as 
$\Phi  = (1- \epsilon) \id + \epsilon \Gamma$ with $\Gamma$ a channel
with Choi rank two unless $\Gamma$ is itself a convex combination
of $\id$ and a unitary conjugation.   As remarked in  \cite{WP},
their results do not apply to maps with Choi rank $> 2$ \cite{SS07},
even for qubits.      Corollary~\ref{cor:d3} implies that a channel 
$$\Phi(\rho) =  (1 -   \epsilon_1 -  \epsilon_2) \rho +  \epsilon_1 X \rho X^\dag + 
\epsilon_2  Z \rho Z^\dag$$
  is neither degradable nor anti-degradable no matter how small
  $  \epsilon_1 +  \epsilon_2$ is.   Thus, there are rank 3 channels with a trit output 
  that are nondegradable, even arbitrarily close to the identity channel.
  
  \medskip

 
 The channel \eqref{bigenv} can be used to make a small observation on
one of the major open questions in quantum information theory, namely,  whether
or not the Holevo capacity, 
\be   \label{Cholv}
    C_{\rm Hv}(\Phi) = \sup_{\pi_k \rho_k} \Big( S \big( \sum_k \pi_k \rho_k \big) - 
          \sum_k \pi_k S(\rho_k) \Big) =  S(\rho_{\rm av}) - Av[S(\rho)]
\ee
is additive under tensor products.    There has been some speculation that
degradabillity of $\Phi$ or, more generally, additivity of the coherent information 
would imply additivity for \eqref{Cholv}.    That this implication need not hold can be 
demonstrated using the channel  \eqref{bigenv}.   First, note
\cite{FW,Storm} that if $\Phi = \Phi_1 \op \Phi_2$, then  
$C_{\rm Hv}(\Phi) =   C_{\rm Hv}(\Phi_1) + C_{\rm Hv}(\Phi_2)$.    For the degradable
channel \eqref{bigenv} this becomes  $C_{\rm Hv}(\Phi) = C_{\rm Hv}({\cal N}) + \log d$,
and it follows that $C_{\rm Hv}(\Phi)$ is additive if and only if  $C_{\rm Hv}({\cal N})$
is additive.    Thus, if a counter-example to additivity for $C_{\rm Hv}({\cal N})$ can
be found, then $C_{\rm Hv}(\Phi)$ would be superadditive despite the fact that it is
degradable.

  \bigskip
 
\noindent{\bf Acknowledgment:}  It is a pleasure for MBR to acknowledge that
this work had its genesis in stimulating discussions with M.M. Wolf during
a workshop in June, 2006 at the ICTP in Trieste, Italy.   GS is
grateful to Debbie Leung, John Smolin, and Jon Yard for many
discussions about degradability. TSC would like to thank Andreas
Winter for much the same reason. The authors also benefitted
from discussions during workshops at BIRS in Banff, Canada in February, 2007 in
in Benasque, Spain in June, 2007 and at the Lorentz Center in Leiden in July, 2007 as well as comments by A. Holevo on an 
earlier draft.  Finally, we are indebted to Michael Nathanson and Yuan Shen for their help in producing Figures~1 and~3, respectively.

The work of MBR was partially supported by the
National Science Foundation under Grants DMS-0314228 and 
DMS-0604900.   The work of GS was partially supported by the UK Engineering and Physical Sciences Research Council.  TC was  
supported by the European Commission, project ``QAP''.

   \bigskip

\appendix

\section{Background}

  \subsection{Arveson commutant lifting theorem}   \label{appA}

The  complement of a channel  $\Phi: M_{d_A} \mapsto M_{d_B}$ is
closely related to a map from  $\Upsilon: M_{d_A}$ to $ M_{d_E}$
defined earlier in
greater generality by Arveson\cite{Arv}.    We explain this following
the notation in Appendix~A of \cite{KMNR}, where it was observed
that the ancilla representation \eqref{rep.anc} is a special case of
Stinespring's fundamental representation theorem \cite{Arv,Paul,Sti}.
For CPT maps, it is more convenient to write this for the dual
$  \wh{ \Phi}:  M_{d_B} \mapsto M_{d_A}$  which is unital
and defined by the relation
$\tr [\wh{\Phi}(X)]^\dag \gamma = \tr X^\dag \Phi(\gamma)$.
The  Stinespring representation then has the form
\be  \label{rep.S}
       \wh{ \Phi}(Q) = V^{\dag} \pi(Q) V
\ee
where
$\pi$ is a representation of the algebra, and $V^\dag V = I_A$ so that
$V$ is a partial isometry.
Arveson's commutant lifting theorem \cite{Arv} defines a map  $\rho
\mapsto X$
by the relation
\be   \label{eq:arv}
X V = V \rho
\ee
with $X$ in the commutant of  $\pi(M_{d_B})$ (or,  $X$
in the commutant of  $\pi({\mathcal B})$ in the general case
  $\Phi : {\mathcal A}\mapsto {\mathcal B}$ of maps on operator
algebras.)
  Then formally, $\Upsilon_{\Phi}(\rho) = (V \rho V^\dag) (V V^\dag)^
{-1}$.
  For matrix algebras, the inverse above is well-defined
  on $(\ker V^\dag)^\perp$; however, in the general setting it may
  require an unbounded operator affiliated with the algebra $
{\mathcal B}$.

As explained in \cite[Chapter 2] {Paul},
for maps on matrix algebra one can choose the representation
as  $\pi(Q) = Q \ot I_E $.
  Then one can write
$V = \sum_j F_j \ot |j \ket $ as a vector of block matrices with
the blocks $F_k$ the Kraus operators of $\Phi$, and
\eqref{rep.S} reduces to  \eqref{rep.anc}.
In the finite dimensional case with the representation chosen
to have the simple form above,  the matrix
$X$ must then have the form $X = I_B  \ot X_E$ and
$X_E  \Phi^C(I) = \Phi^C(\rho)$.

Since \eqref{eq:arv} implies
$    V Q V^\dag  =   (I_B  \ot X_E) VV^\dag   $,  using the block vector
expression for $V$ above gives
\be    \label{arv2}
  \sum_{jk}  F_j Q F_k^\dag \ot |j\kb  k|
=    \sum_{jk}  F_j  F_k^\dag  \ot X_E |j \kb k|
  \ee
Then taking the partial trace over $B$ and using $\tr F_j F_k^\dag =
\delta_{jk} \tau_k$  yields
\be   \label{arv3}
      \sum_{jk}  ( \tr  F_j Q F_k^\dag ) \, |j \kb k|  & = &  \sum_
{jk}  \tr (F_j  F_k^\dag)  \ot X_E |j \kb k|  \\
          & = & X_E   \Big(\sum_{jk}  \tr (F_j  F_k^\dag)  |j \kb k|
\Big).
          \ee
Since the left side of \eqref{arv3} is  exactly the form of $\Phi^C
(\rho)$
given by Eq. (6) in \cite{KMNR},
we can conclude that
\be   \label{arv4}
    \Phi^C(\rho) = X_E \Phi^C(I_A) =  \wtd{\Upsilon}_{\Phi}(\rho)
\Phi^C(I_A)
\ee
with $ \wtd{\Upsilon}_{\Phi}(\rho) \equiv X_E  = \tfrac{1}{d_B}
\trp_B \, \Upsilon_\Phi(\rho)$
obtained from Arveson's Theorem.

Although this establishes a relation between the complement of a channel
and Arveson's lifting,  it might appear that one can only use \eqref
{arv4} to obtain
Arveson's channel  from the complement, but not the reverse.
However, one can also do the latter by choosing the $F_k$ to be
the eigenvectors of the Choi matrix of $\Phi$ after unstacking
and renormalized so that  $\tr F_k F_k^\dag = \tau_k$ are the non-zero
eigenvalues of the Choi matrix.    Then
\be
  \Phi^C(I_A) = \sum_{k = 0}^{d_E} \tau_k |k\kb k| \equiv D_{\Phi}
\ee
is unitarily equivalent to the projection of the Choi matrix of $\Phi$
onto the orthogonal complement of its kernel.   To see this write
the spectral representation of the Choi matrix
as  $\ds{ \sum_{k = 0}^{d_A d_B} \tau_k |f_k \kb f_k|  }$ with $|f_k
\ket$ the
normalized eigenvectors corresponding to $F_k$.   Omitting the
eigenvectors with $\tau_k = 0$  gives $D_{\Phi}$.
Thus $D_{\Phi}^{-1} = \sum_k \tau_k^{-1} |k\kb k|$ is well defined
and
\be
        \Phi^C(\rho) = \wtd{\Upsilon}_{\Phi}(\rho)  D_{\Phi}
\qquad \hbox{or} \qquad
           \wtd{\Upsilon}_{\Phi}(\rho) =   \Phi^C(\rho) D_{\Phi}^{-1}.
\ee
This allows one to obtain either the complement from Arveson's
channel or
Arveson's channel from the complement.

\subsection{Degradability implies additivity}  \label{appB}

The standard definition of the coherent information of a 
channel  $\Phi: {\mathcal B}( {\mathcal H}_{A_2}) \mapsto  {\mathcal B}( {\mathcal H}_{B})$
 with respect to a reference state $\rho$  is
\be   \label{cohpur}
I^{\rm coh}(\Phi,\rho) = S[ \Phi(\rho)] - S[(I \ot \Phi)(\proj{\chi}) ] 
\ee
with $| \chi \ket$  in 
$ {\mathcal H}_{A_1A_2} \equiv {\mathcal H}_{A_1} \ot  {\mathcal H}_{A_2}$ satisfying
the purification condition $  \trp_{A_1}  \proj{\chi} = \rho$.   But by the
Stinespring representation
\be
     ( I \ot \Phi)(\proj{\chi}) =  \trp_{E} \,(I \ot  V) \proj{\chi} (I \ot V)^\dag    
\ee
with $V:   {\mathcal H}_{A_2 E} \mapsto  {\mathcal H}_{B E} $ a partial isometry.
Now, since $(I \ot  V) \proj{\chi} (I \ot V)^\dag$ is a pure state,
\be
    S[(I \ot \Phi)(\proj{\chi}) ]  & = &  S\big[  \trp_{E} \,(I \ot  V) \proj{\chi} (I \ot V)^\dag   \big] \nn  \\
       & = &  S\big(  \trp_{A_1 B} \,(I \ot  V) \proj{\chi} (I \ot V)^\dag   \big)  \nn \\
       & = & S ( \trp_{B}   V  \rho    V^\dag ) = S\big[ \Phi^C(\rho) \big]
\ee 
Inserting this in \eqref{cohpur} yields \eqref{cohref}.

To show that degradability implies additivity, observe that the
monotonicity of relative entropy  $H(\rho, \gamma) \equiv \tr \rho (\log \rho - \log \gamma)$
under CPT maps implies 
 \bee   
\lefteqn{    - S[\Phi^C\! (\rho_{AB})] + S[\Phi^C\! (\rho_A)] + S[\Phi^C\! (\rho_B)] } \qquad \qquad \\
    & = & H\big[ (\Phi^C \ot \Phi^C)(\rho_{AB}),(\Phi^C \ot \Phi^C)( \rho_A \ot \rho_B )\big ]  \\
    & = & H\big[( \Psi \ot \Psi ) \circ (\Phi \ot \Phi)(\rho_{AB}), 
              ( \Psi \ot \Psi ) \circ (\Phi \ot \Phi) (\rho_A \ot \rho_B)\big]  \\
    & \leq & H\big[(\Phi \ot \Phi)(\rho_{AB}), 
             (\Phi \ot \Phi) (\rho_A \ot \rho_B)\big] \\  
        & = &   -S[\Phi(\rho_{AB})] + S[\Phi(\rho_A)] + S[\Phi(\rho_B)].
        \eee
        Rearranging gives
  $$ S[\Phi(\rho_{AB})]  \mm   S[ \Phi^C\! (\rho_{AB}) ] \,  \leq \, 
   S[\Phi(\rho_A)]  \mm  S[\Phi^C\! (\rho_A)] \,  + \,  S[\Phi(\rho_B)]  \mm S[\Phi^C\! (\rho_B)] $$
which by  \eqref{cohref}  is equivalent to
\be
   I^{\rm coh}(\Phi \ot \Phi,\rho_{AB})  \leq   I^{\rm coh}(\Phi ,\rho_{A}) +  I^{\rm coh}(\Phi ,\rho_{B}).
\ee
 This implies   $ I^{\rm coh}(\Phi \ot \Phi)  \leq  2  I^{\rm coh}(\Phi)$ and the reverse
 inequality is trivial.  This argument clearly extends to tensor products of 
 different degradable channels $ \Phi_1 \ot \Phi_2$ and hence implies  
 $ I^{\rm coh}(\Phi^{ \ot m})  = m  I^{\rm coh}(\Phi)$.

\subsection{Properties of antidegradable channels}
\label{appendix:Convex}

In this section,  we show that the set of antidegradable channels is
convex.
To do this, we first prove another result that is of independent
interest.
\begin{lemma}   \label{lemm:anti}
Let $\Gamma : \cB(\cH_A)\mapsto \cB(\cH_B)$ be an anti-degradable CPT
map
and $\Delta : \cB(\cH_B)\mapsto \cB(\cH_C)$ any CPT map.  Then  the
channel
$\Delta \circ \Gamma$ is also anti-degradable.
\end{lemma}
\pf  Let  $\cH_E$ and $\cH_D$ be the environments for
$\Gamma$ and $\Delta$ respectively and let
$U: \cH_A \mapsto \cH_{BE}$  and  $V: \cH_B \mapsto \cH_{CD}$
denote the corresponding partial isometries for their
 Stinespring representations as in  \eqref{rep.anc}.
Then the complement of  $\Delta \circ \Gamma$
maps  $\cB(\cH_A)\mapsto \cB(\cH_{DE}) $ and satisfies
 \begin{equation}
(\Delta \circ \Gamma)^C (\rho) = \trp_{C}  \, (V \otimes I_E) U\rho U^
\dag (V^\dag \otimes I_E).
\end{equation}
Furthermore, since the range of $V$ is $ \cH_{CD}$ and $V^\dag V = I_B $
\be
\trp_D \, (\Delta \circ \Gamma)^C (\rho)  & = &
 \trp_{CD}  \, U\rho U^\dag (V^\dag \otimes I_E) (V \otimes I_E) \nn \\
    & = & \trp_C \, U\rho U^\dag ~ =  ~ \Gamma^C(\rho).
\ee
By assumption, there is a channel $ \Lambda :  \cH_E \mapsto \cH_B $
such that
$ \Lambda \circ \Gamma^C =  \Gamma$.  But then
\begin{equation}
( \Delta \circ \Lambda  \circ \trp_D) (\Delta \circ \Gamma)^C (\rho)
= \Delta \circ \Gamma(\rho),
\end{equation}
which implies  that $\Delta\circ \Gamma$ is antidegradable.  \qquad \qed

\begin{thm}
The set of anti-degradable channels is convex.
\end{thm}
\pf  Let $\Phi_0$ and $\Phi_1$ be antidegradable channels and
consider the channel
\begin{equation}
\Gamma = (1-p)\Phi_0 \otimes |0\rangle \langle 0 |_F + p \Phi_1
\otimes |1\rangle \langle 1 |_F,
\end{equation}
whose complement is
\begin{equation}
\Gamma^C = (1-p)\Phi_0^C \otimes |0\rangle \langle 0 |_G + p \Phi_1^C
\otimes |1\rangle \langle 1 |_G.
\end{equation}
By assumption,  there exist   $\Psi_j$ such that $ \Psi_j \circ
\Phi_j^C = \Phi_j .$
With the Kraus operators of $\Psi_j$ denoted $\{A^j_k\}_k$, define
\begin{equation}
A_{j} = A^0_k \otimes |0\rangle \langle 0| + A^1_k \otimes |1\rangle
\langle 1|,
\end{equation}
and let  $\Psi$ be the channel with Kraus operators $A_k$.  Then
\begin{eqnarray}
\Psi \circ \Gamma^C &=& (1-p)(\Psi_0 \circ \Phi_0^C) \otimes |0
\rangle \langle 0| + p(\Psi_1 \circ \Phi_1^C) \otimes |1\rangle
\langle 1|   \nn \\
 & = & (1-p)\Phi_0 \circ \otimes |0\rangle \langle 0| + p\Phi_1
\circ \otimes |1\rangle \langle 1|
  = \Gamma
\end{eqnarray}
so that $\Gamma$ is antidegradable.   Then applying Lemma~\ref
{lemm:anti} with $\Delta = \trp_F$
implies that the channel
\begin{equation}
\trp_F  \, (1-p)\Phi_0 \otimes |0\rangle \langle 0 |_F + p \Phi_1
\otimes |1\rangle \langle 1 |_F) =
   (1-p)\Phi_0 + p\Phi_1.
\end{equation}
 is anti-degradable.  This proves that the convex combination
 $\Phi = (1-p)\Phi_0 + p\Phi_1$ is anti-degradable.  \qquad \qed

\section{Qubit channels}  \label{app:qubit}

\subsection{Qubit channel representations and conditions}  \label{app:qbasic}

We first recall some well-known facts about qubit channels
from \cite{KR} and \cite{RSW}.
A linear map $\Phi:M_2 \mapsto M_2$ can be represented by the matrix
$T_{\Phi}$ with elements $ \tr \sigma_j \Phi(\sigma_k)$.    When $\Phi$
has the form
\be  \label{qmap}
    \Phi:    \half [I + \sum_j w_k \sigma_k ] \longmapsto 
        \half [I + \sum_k (t_k + \lambda_k w_k) \sigma_k)]
\ee
this matrix  is
\be   \label{tphi}
T_\Phi  =  \pmx   1 & 0 & 0 & 0 \\  t_1 & \lambda_1 & 0 & 0 \\
                                 t_2 & 0 & \lambda_2 &  0 \\  t_3 &  0
& 0 &  \lambda_3  \emx.
                                 \ee
 It was shown in   \cite{RSW} that when $t_1 = t_2 = 0$ a linear
 map of the form \eqref{qmap} is completely positive (CP) if and only if
 all $| \lambda_k| \leq 1$ and  \be   \label{CP}
    (\lambda_1 \pm \lambda_2)^2 \leq ( 1 \pm \lambda_3)^2 - t_3^2,
 \ee                         and that the map has Choi rank $d_E \leq
2$,  if and  only if equality holds in \eqref{CP}.    In that case,
\be  \label{extcond}
\lambda_3 = \lambda_1 \lambda_2,  \quad \hbox{and}  \quad
t_3^2 = (1 - \lambda_1^2)(1 - \lambda_2^2).
\ee  
Channels satisfying \eqref{extcond} can  can be represented by the
matrix
\be   \label{ext}
T_\Phi  =  \pmx   1 & 0 & 0 & \\  0 & \cos u & 0 & 0 \\
                                 0 & 0 & \cos v &  0 \\  \sin u \sin v
&  0 & 0 &  \cos u \cos v  \emx
                                 \ee
with $u = \cos^{-1}(\lambda_1), v = \cos^{-1}(\lambda_1)$.

As noted after Theorem~\ref{thm:CNG},  up to unitary conjugations, 
unital qubit maps  can be written as
$\Phi(\rho) = \ds{ \sum_{k=0}^3 a_k^2 \, \sigma_k \rho\,  \sigma_k }$
with  $\sum_k a_k^2 = 1$ and the convention  $\sigma_0 = I $. The
matrix representative \eqref{tphi} has $t_i=0$ and
\be
\lambda_k = a_0^2 + a_k^2 -
a_i^2 - a_j^2
\ee
with $i,j,k$ distinct,  or, equivalently,
\be    \label{a_lam}
     \pmx   1  \\ \lambda_1 \\   \lambda_2 \\    \lambda_3  \emx  =
                               \pmx 1 & ~ \, 1 & ~ \, 1 & ~ \, 1 \\
1 & ~ \, 1 & -1 & -1 \\
                                1 & -1 & ~ \, 1 & -1  \\ 1 & -1 & -1 &
~ \, 1  \emx                                  \pmx a_0^2 \\  a_1^2  \\
a_2^2  \\ a_3^2  \emx.
\ee

\subsection{Proof of Theorem~\ref{thm:2to2}}  \label{app:qdeg}

We now present a proof of Theorem~\ref{thm:2to2} different
from that in \cite{WP2}.
 Denote a channel parameterized as in \eqref{ext} by
   $\Phi(u,v)$.   The amplitude-damping channels are those with $u = v$
   and satisfy  $\Phi(u_1, u_1) \circ \Phi(u_2,u_2) = \Phi(u_3,u_3) $ with
   $u_3 = \cos^{-1}( \cos u_1 \cos u_2 )$.   However, it is {\em not}
     true in general that
   $  \Phi(u_1, v_1)  \circ \Phi(u_2,v_2) = \Phi(u_3,v_3) $.   The next
   theorem shows that this holds in a very special case.

\begin{thm} \label{qub}
Let  $\Phi(u,v)$ be a qubit channel of the form \eqref{ext}.   Then

a)  $\Phi^C(u,v) =  \Phi(v - \tfrac{\pi}{2} ,u - \tfrac{\pi}{2}) $, and

b) if $|\sin v| \leq |\cos u|$, $     \Phi(\theta_1, \theta_2) \circ \Phi(u,v) = \Phi(v -
\tfrac{\pi}{2} ,u - \tfrac{\pi}{2})$
with
   \be   \label{theta}
    \theta_1 = \cos^{-1}(\sin v/ \cos u)  \qquad  \theta_2 =
\cos^{-1}(\sin u/ \cos v).
    \ee    \end{thm}
 Combining Theorem~\ref{qub}
  with the fact  that the Kraus operators for \eqref{ext}
 are  $A_+$ and $A_-$ defined in \eqref{qextK}  yields
   Theorem~\ref{thm:2to2}.           Note that part (a) implies that a
simple algorithm to map
   $\Phi \leftrightarrow \Phi^C$ is to change
$\cos u   \leftrightarrow \sin v$ and $\cos v   \leftrightarrow \sin u$.
(It is important that one change {\em both}  $\sin \leftrightarrow \cos$
{\em and}  $u \leftrightarrow v$.)

\noindent{\bf Proof of Theorem~\ref{qub}:} To prove (a), we begin with
the fact \cite{RSW}
that the Kraus operators for \eqref{ext}
are  $A_+$ and $A_-$ defined in \eqref{qextK} which we write in the
in the compact form
   \bee
   F_1 = A_+ =  aI + b \sigma_z   \qquad          F_2  = A_- =
c\sigma_x + i d \sigma_y.    \eee Next, we use the
observation \cite[Eq. (6)]{KMNR}
   that if $\Phi(\rho) = \sum_k F_k \rho F_k^\dag$
   then $\Phi^C(\rho) $ is the matrix with elements  $\tr F_j \rho
F_k^\dag$.  Then for $\rho = \half \big[I + \sum_j w_j \sigma_j\big]$ a
straightforward computation gives
  \begin{eqnarray}
     \Phi^C(\rho) & = &   \pmx  a^2 + b^2 + 2 ab \, w_3  &  (ac+bd)
w_1 -i (bc + ad)\, w_2 \\
           (ac+bd) w_1 + i (bc + ad)\, w_2 & c^2 + d^2 - 2 ab \, w_3
\emx   \\   \nn
       & =  &  \half \big[  I +  \sin v \,  w_1 \sigma_1 + \sin u \,
w_2 \sigma_2 +                (\cos u \cos v + \sin u \sin v \,  w_3)
\sigma_3 \big] , \quad \qquad
  \end{eqnarray}
which establishes part (a).
       
    To prove part (b), we rewrite this in the form \eqref{ext}
and  compute $\Phi^C \circ \Phi^{-1}$ to get
  \be   \label{d}
     \Psi & = &   \pmx   1 & 0 & 0 & 0\\  0 & \sin v & 0 & 0 \\
                                 0 & 0 & \sin u &  0 \\  \cos u \cos v
&  0 & 0 &  \sin u \sin v  \emx
          \pmx   1 & 0 & 0 & 0\\  0 & \frac{1}{\cos u} & 0 & 0 \\
                                 0 & 0 &  \frac{1}{\cos v} &  0 \\
               - \frac{sin u \sin v}{ \cos u \cos v} &  0 & 0 &
\frac{1}{ \cos u \cos v } \emx  \nn \\
     & = &      \pmx   1 & 0 & 0 & 0\\  0 & \frac{\sin v}{\cos u} & 0 & 0 \\
                                 0 & 0 &  \frac{\sin u}{\cos v} &  0
\\                   \cos u \cos v - \frac{sin^2 u \sin^2 v}{ \cos u
\cos v} &  0 & 0 &  \frac{sin u \sin v}{ \cos u \cos v }
  \emx    .      \ee     To see if  $\Psi$ is CPT, we first
apply the necessary condition   $ |\lambda_j|   \leq 1$ for $ j = 1,2,3$
  to \eqref{d}.
Since  $|\sin v | \leq |\cos u|  \Leftrightarrow  |\sin u | \leq |\cos
v| $ this condition
is either satsified for all $\lambda_j$ or for none (if it is none, the
map will be anti-degradable.)   Then it suffices to see if    \eqref{CP}  holds.
 \be
    t_3^2 & = &  \frac{1}{ \cos^2 u \, \cos^2 v} \big(  \cos^2 u \,
\cos^2 v - \sin^2 u \, \sin^2 v \big)^2  \nn \\
    & = &  \frac{1}{ \cos^2 u \, \cos^2 v} \big( \cos^2 u \, (1 -
\sin^2 v) - (1 - \cos^2 u) \sin^2 v \big)^2  \nn \\
    & = &  \frac{1}{ \cos^2 u \, \cos^2 v} \big(  \cos^2 u - \sin^2 v
\big)^2 \nn \\
    & = &  \frac{1}{ \cos^2 u \,  \cos^2 v} ( \cos^2 u - \sin^2 v) (
\cos^2 v - \sin^2 u)  \nn \\
    & = &   \big( 1 - \frac{ \sin^2 v}{\cos^2 u} \big)   \big( 1 -
\frac{ \sin^2 u}{\cos^2 v} \big)
           ~ = ~ (1- \lambda_1^2)(1- \lambda_2^2) .
\ee
Thus, $\Psi$ is not only CP, it is also a  map of the form  \eqref{ext}
with  $\lambda_1 = \cos  \theta_1 = \frac{\sin v}{\cos u}$ and
$\lambda_2 =  \cos  \theta_2 = \frac{\sin u}{\cos v}$, or equivalently
 $   \Phi(\theta_1, \theta_2) $ with $\theta_j$ given by \eqref{theta}.  Thus
 \eqref{d} becomes  $  \Phi(\theta_1, \theta_2) =    \Phi(v -
\tfrac{\pi}{2} ,u - \tfrac{\pi}{2}) \circ \Phi^{-1}(u,v)$
 which implies part (b).  \quad \qed

\subsection{Anti-degradable unital qubit channels}  \label{app:qanti}


Since the Kraus operators
for a unital qubit channel are  $a_k \sigma_k$, it follows from
\cite[Eq. (4.2)]{KMNR}  that  
\be    
\Phi^C(\rho) = \pmx a_0^2 & a_0a_1 & a_0a_2 & a_0a_3 \\
                        a_1a_0 & a_1^2 & a_1a_2 & a_1a_3 \\
                        a_2 a_0 & a_2a_1 & a_2^2 & a_2a_3 \\
                        a_3 a_0 & a_3a_1 & a_3 a_2 & a_3^2 \emx    ~ * ~
                        \pmx w_0 & w_1 & w_2 & w_3 \\
                             w_1 & w_0 & -i w_3 & i w_2 \\
                             w_2 & i w_3 & w_0 & -i w_1 \\
                             w_3 & -iw_2 & i w_1 & w_0 \emx,
\ee
where $*$ denotes the pointwise Hadamard product.
$\Phi^C$ can be represented by the $16 \times 4$ matrix
with elements  $  \tr |e_j\kb e_k | \Phi^C(\sigma_m)   $
\be   \pmx  a_0^2 & 0 & 0 & 0 \\
                 0 & a_0a_1 & 0 & 0 \\
                 0 & 0 & a_0a_2 & 0 \\
                 0 & 0 & 0 & a_0a_3 \\ ~~ \\
                 0 & a_1a_0 & 0 & 0 \\
                 a_1^2 & 0 & 0 & 0  \\
                  0 & 0 & 0 & -i \, a_1a_2 \\
                  0 & 0 &  i \, a_1a_3 & 0\\  ~~ \\
                   0 & 0 & a_2 a_0 & 0 \\
                    0 & 0 & 0 & i \,  a_2 a_1 \\
                    a_2^2 &  0 & 0 & 0 \\
                    0  & - i \,  a_2 a_3 &  0 & 0 \\ ~~ \\
                    0 & 0 & 0 &  a_3a_0 \\
                     0 & 0 &  -i \,  a_3 a_1 & 0 \\
                     0 & i \,  a_3 a_2 &  0 & 0  \\
                     a_3^2 &  0 & 0 & 0 \\
      \emx .   \ee
If  $\Phi$ is anti-degradable, i.e.\ $\Psi \circ \Phi^C = \Phi$,  the map
$\Psi$ can be represented by a $4 \times 16$ matrix with elements
$\tr \sigma_n  \Psi( |e_j\kb e_k | )   $
\be   \label{Psi}  \pmx
   1 & 0 & 0 & 0 & \, & 0 & 1 & 0 & 0 & \, & 0& 0 & 1 & 0 & \, & 0& 0
& 0 & 1 \\
     0 & x_1 & 0 & 0 & \, & x_1 & 0 & 0 & 0 & \, & 0& 0 & 0 & iy_1 &
\, & 0& 0 & \! -iy_1 & 0\\
       0 & 0 & x_2 & 0 & \, & 0 & 0 & 0 & \! -iy_2 & \, & x_2& 0 & 0 &
0 & \, & 0& iy_2 & 0 & 0\\
         0 & 0 & 0 & x_3 & \, & 0 & 0 & iy_3 & 0 & \, & 0& \!  -iy_3 &
0 & 0 & \, & x_3 & 0 & 0 & 0
\emx \ee where $x_k$ and $y_k$ must be chosen to satisfy
\be   \label{cond}
   2 a_0 a_k  \, x_k + 2 a_i a_j \, y_k =   a_0^2 + a_k^2 -  a_i^2  -
a_j^2  = \lambda_k
\ee
with $i,j,k$ distinct.
Although there are many solutions for $x_k, y_k$, only those which
yield a CP map are
acceptable.   To check this, one needs to find the Choi matrix of $\Psi$.
Each column of \eqref{Psi} defines one of the blocks in the Pauli basis,
e.g.,  the block in the row 1 and col 3 is $x_2 \sigma_y $. Thus the full Choi
matrix for $\Psi$ is
\be   \label{choi} \pmx
     1 & 0 & \, & 0 & x_1 & \, & 0 & -ix_2 & \, &    x_3 & 0 \\
     0 & 1 & \, &x_1  & 0 & \, & i x_2  & 0 &\, &  0 & - x_3 \\  \,\, \\
     0 &  x_1 & \, & 1 & 0 & \, & iy_3 & 0 & \, & 0 & -y_2 \\
     x_1  & 0 & \, &  0 & 1  & \, & 0 & -iy_3 & \, & y_2  & 0 \\  \,\, \\
     0 &-ix_2  & \, &  -i y_3 & 0 & \, &  1 & 0 & \, & 0 & i y_1 \\
    ix_2 & 0 & \, &  0 & i y_3 & \, &    0 & 1 & \, &  i y_1 & 0 \\  \,\, \\
    x_3 & 0 & \, &  0 &y_2 & \, &  0 & - i y_1 & \, & 1 & 0  \\
     0 &  -x_3 & \,  &  - y_2 & 0       & \, &  - iy_1  & 0 & \, & 0 & 1
\emx .\ee
By conjugating with a suitable permutaton matrix, one can see
that this contains two blocks, both unitarily equivalent to the matrices
  \be   \label{antimat}   \wtd{Y} = \pmx
 1 &  x_1 &    x_2 &  x_3  \\
x_1  & 1  &  y_3 & y_2  \\     x_2  &   y_3 & 1 &   y_1 \\     x_3 &
y_2 &    y_1& 1 
\emx   \qquad \qquad
 Y = \pmx
 1 &  x_3 &    x_1 &  x_2  \\
x_3  & 1  &  y_2 & y_1  \\     x_1  &   y_2 & 1 &   y_3 \\     x_2 &
y_1 &    y_3 & 1 
\emx .\ee
The matrix  $\wtd{Y}$  is, up to phase factors, embedded in \eqref{choi};
an additional permutation yields the unitarily equivalent matrix $Y$,
which we prefer to use. 
 Thus, \eqref{choi} is positive semi-definite if and only if
\eqref{antimat} is,  which
requires
\be  \label{cond1}
    |x_k| \leq 1  \qquad  \hbox{and}   \qquad    |y_k| \leq 1 .\ee
When  \eqref{cond1} holds, \eqref{antimat}  is  positive
semi-defininite if and only if
\be    \label{pos}
& ~ &     \pmx    x_1 &  x_2  \\   y_2 & y_1  \emx
   \pmx   1 &   y_3 \\    y_3& 1 \emx^{-1}       \pmx    x_1 &  y_2
\\   x_2 & y_1  \emx     \leq         \pmx   1 &   x_3 \\    x_3 & 1
\emx         .\ee
       Using        $\big( \begin{smallmatrix}  1 &   ~y  \\    ~y & 1
\end{smallmatrix} \big)^{-1}  = \tfrac{1}{1-y^2}
         \big( \begin{smallmatrix}  1 &   -y  \\    -y & 1
\end{smallmatrix} \big) $ this is straightforward
         to evaluate.
          In some cases,  conjugating with the
Hadamard gate         $H = 2^{-1/2}   \big( \begin{smallmatrix}  1 &   ~
1   \\    1 & -1 \end{smallmatrix} \big) $
  gives a more useful expression.  In particular,
\begin{itemize}  
 \item When  $x_1 = x_3$ and  $y_1 = y_3$, 
   \eqref{pos}  becomes
            \be   \label{pos2}
 \tfrac{1}{1+y_3}  \pmx (x_1+y_1)^2 &  x_1^2 - y_1^2 \\   x_1^2 -
y_1^2 & (x_1-y_1)^2 \emx  \leq       \pmx 1+x_3 & 0 \\ 0 & 1-x_3 \emx
\ee    or, equivalently,
\be \label{pos_sig}
  ( 1 + y_3 - x_1^2 - y_1^2) I - (x_1^2 - y_1^2) \sigma_3 +
[x_3(1+y_3) + 2x_1 y_1] \sigma_3 \geq 0.
\ee

\item  When $x_k = y_k$, the matrix$ (H \ot I) Y (H \ot I) $
is precisely the Choi matrix of the unital map    
\be
 \half   [I + \sum_j x_k \sigma_k ] \longmapsto  \half [I + \sum_k  
\lambda_k x_k \sigma_k] .
\ee
Thus, $\Psi$ is CP if and only if $|x_k| \leq 1$ and
$(x_1 \pm x_2)^2 \leq (1 \pm x_3)^2 $.   This can also be
seen by obvserving that conjugating both sides of \eqref{pos} with $H$ 
yields diagonal matrices satisfying
\bee
   (x_1 I + x_2 \sigma_3)(I + x_3 \sigma_3)^{-1}  (x_1 I + x_2 \sigma_3)
        \leq (I + x_3 \sigma_3).
\eee

\end{itemize}

It should be pointed out that   \eqref{Psi} is not the most general possible
degrading map.    For example,  one could change its first row to \bee  \pmx
      1 & t & 0 & 0 & \, & -t & 1 & 0 & 0 & \, & 0& 0 & 1 & 0 & \, &
0& 0 & 0 & 1
\emx  \eee
 This will not affect \eqref{cond}, but it will introduce
non-zero cross-terms in the block structure used to reduce the
positivity of \eqref{choi} to that of \eqref{pos}.    The positivity of
 \eqref{pos}  will still be necessary, but  the cross-terms  will
introduce additional constraints
without relaxing any other requirements.    Thus, there is no loss of
generality in assuming that the degrading map has the form $\Psi$.

\subsection{Anti-degradable channels with one $a_k = 0$}
\label{app:ant2paul}

For notational simplicity, we assume $a_0 = 0$ and
$a_j \neq 0$ for $j = 1,2,3$.  Then we can assume $x_k = 0$ and, with
$i,j,k$ distinct
\be   \label{2pcond1}
    y_k = \frac{2 a_k^2 - 1}{2a_ia_j} = \frac{1 - 2 a_i^2 - 2 a_j^2
}{2 a_i a_j}
\ee
When $x_k = 0$, \eqref{pos} becomes
\be   \label{2pcond}
  y_1^2 + y_2^2 + y_3^2 - 2 y_1 y_2 y_3 \leq 1
\ee
which is equivalent to the condition that the $3 \times 3$
subdeterminant of \eqref{antimat} is $\geq 0$.   

When   $a_1^2 = a_2^2,  a_3^2 - 1 = 2a_1^2$, and 
$y_1 =  y_2 = \frac{1 a^2 -1 }{2a \sqrt{1-2 a^2}},
y_3 = \frac{1-4a^2}{2a^2}$, the condition \eqref{2pcond}
is equivalent to $2 y_1^2 (1 - y_3) \leq 1 - y_3^2$ so that
for $y_3 \neq \pm 1$, \eqref{2pcond} is equivalent to
$2 y_1^2  \leq 1+y_3$.   But  since 
$2 y_1^2 =  \frac{1 - 2a^2 }{2a^2} = 1+y_3$, \eqref{2pcond} always holds
with equality when we choose $y_1 = y_2$.  Moreover, the
choice, $y_1 =  - y_2$ gives a stronger condition when $y_3 > 0$,
 but does not yield additional solutions.

In the general case $a_1 \neq a_2$, substituting \eqref{2pcond1} into  
\eqref{2pcond} gives
\be  \label{a0cond1}
 4 a_1^2 a_2^2 a_3^2  \geq  a_1^2(2a_1^2 \mm 1)^2 + a_2^2(2a_2^2 \mm 1)^2
   + a_3^2(2a_3^2 \mm  1)^2 -  (2a_1^2 \mm  1)(2a_2^2 \mm  1)(2a_3^2 \mm  1).
\ee
By using $a_k^2 = 1 - a_i^2 - a_j^2$
this can be reduced to an inequality in two variables, which,
perhaps surprisingly, can also be shown to hold with equality
after some rather tedious algebra.
   Thus, in the situation considered here with the choices above
 for $y_k$, the matrix \eqref{pos} is positive semi-definite if all
 $y_k^2 \leq 1$.  
 
 The condition $|y_3|^2  \leq 1$   becomes
\be   \label{2pcond2}
   ( 1 - 2 a_1^2 - 2a_2^2)^2 \leq 4 a_1^2 a_2^2
\ee
After inverting \eqref{a_lam} and substituting, one finds
\be   \label{2plam}
    4 \lambda_3^2 \leq  (1 - \lambda_3)^2 - (\lambda_1 - \lambda_2)^2
\ee
or, equivalently, 
 \be    \label{2pcond3}
    (\lambda_1 - \lambda_2)^2 \leq (1 - 3 \lambda_3 )(1 + \lambda_3)
\ee
which implies $-1 \leq \lambda_3 \leq \frac{1}{3} $.  The conditions
for $k = 1,2$ are equivalent.    Thus, a necessary and sufficient
condition that a channel with  $a_0 = 0$ is anti-degradable is  
 \be    \label{2pcondijk}
    (\lambda_i - \lambda_j)^2 \leq (1 - 3 \lambda_k)(1 + \lambda_k)
\ee
for any permutation   $\{i,j,k \}$ of $\{1,2,3\}$.    Similar conditions hold
if  $a_n^2 = 0$ for $n =1,2,3$ and $a_n$ replaced by $a_0$.

Recall that in a fixed basis, the unital qubit maps correspond to a tetrahedron
 with the multiplier  $ [\lambda_1, \lambda_2,\lambda_3] $
defining a point in in ${\bf R}_3$.
Each condition  $a_n^2 = 0$ describes a  triangular ``face'' of this
 tetrahedron.    In particular, the face with $a_0^2 = 0$ is the convex 
 hull of the 3 points  
$$ [1,-1,-1],  \quad [-1,+1,-1],  \quad  [-1,-1,1]  $$
corresponding to conjugation with $\sigma_k$ for $k = 1,2,3$
respectively.       See Fig.~1.

\begin{itemize}

\item Each of the  edges of the tetrahedron and, hence, the edges
of the face correspond to degradable channels, with only
the midpoints anti-degradable as well as degradable.  

\item  The EB maps correspond to triangles whose vertices
are  midpoints of the edges of the face, i.e., maps whose
multipliers are permutations of $[0,0,\pm1]$.

\item  The boundary of the anti-degradable region is described by
 curves  obtained as the intersection of the surface
of points for which equality holds in \eqref{2pcondijk} with a face.
Projected onto one of the faces, these curves form a circle.

  \end {itemize} 
  
  \begin{figure}[h]

 \begin{center}
  \includegraphics*[width=7cm,height=7cm,keepaspectratio=true]{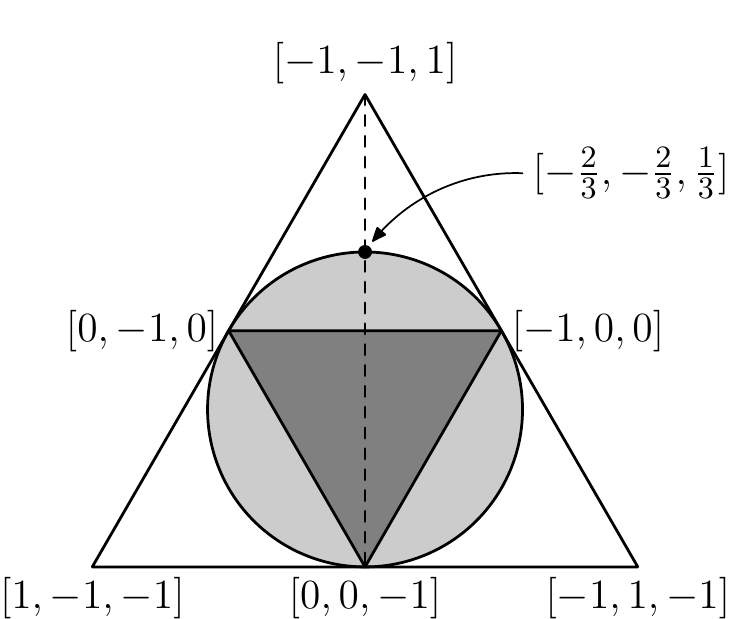}   ~~~   
  \end{center}

\caption{The face of the tetrahedron of unital qubit maps with $a_0^2 = 0$.   
The EB region is the small darkly shaded triangle; the anti-degradable 
region is the circle and its  interior.
       The dashed line corresponds to channels
 with multiplier  $[-x,-x,2x-1]$  unitarily equivalent to two-Pauli channels,
 with the extreme anti-degradable one marked with a dot.}
    \label{fig:face}

\end{figure}

The so-called ``two-Pauli'' channel has (up to permutations of $1,2,3$),
$a_3 = 0$,  $a_1 = a_2 = t,  a_0 = \sqrt{1 - 2t^2}$ with $0 < t^2 \leq
\half$ and multiplier $[1-2t^2, 1-2t^2, 1-4t^2]$.  
Switching $a_0 \leftrightarrow a_3$ does not change the 
analysis above  in any essential way; it suffices to set $y_3 = 0 $ and replace 
$y_3$ by $x_3$ in \eqref{2pcond1} and what follows.   Moreover, this change
does not  affect   \eqref{2pcond2}
which  becomes $|1 - 4 t^2 | \leq 2 t^2$.    Thus, we can conclude that a
two-Paul channel is anti-degradable if and only
if   $\frac{1}{6} \leq t^2 \leq \half$.    This is larger than the
entanglement breaking range  $\frac{1}{4} \leq t^2 \leq \half$, and thus
gives (after including permutations and conjugations) 12 new extreme 
points of the anti-degradable channels with $t^2 = \frac{1}{6}$, e.g., 
corresponding to multiplier $[ \frac{2}{3},  \frac{2}{3},  \frac{1}{3}]$. 
Conjugating this with $\sigma_3$  gives a family of channels with multipliers
of the form $[-x,-x, 2x - 1]$ with $0 \leq x = 1 - 2t^2 \leq 1$ corresponding  
to the dashed line shown in Figure~\ref{fig:face}.

\subsection{Anti-degradable depolarizing channel}    \label{app:antdep}

\begin{figure}[h] 
\centerline{  \includegraphics[height=8cm]{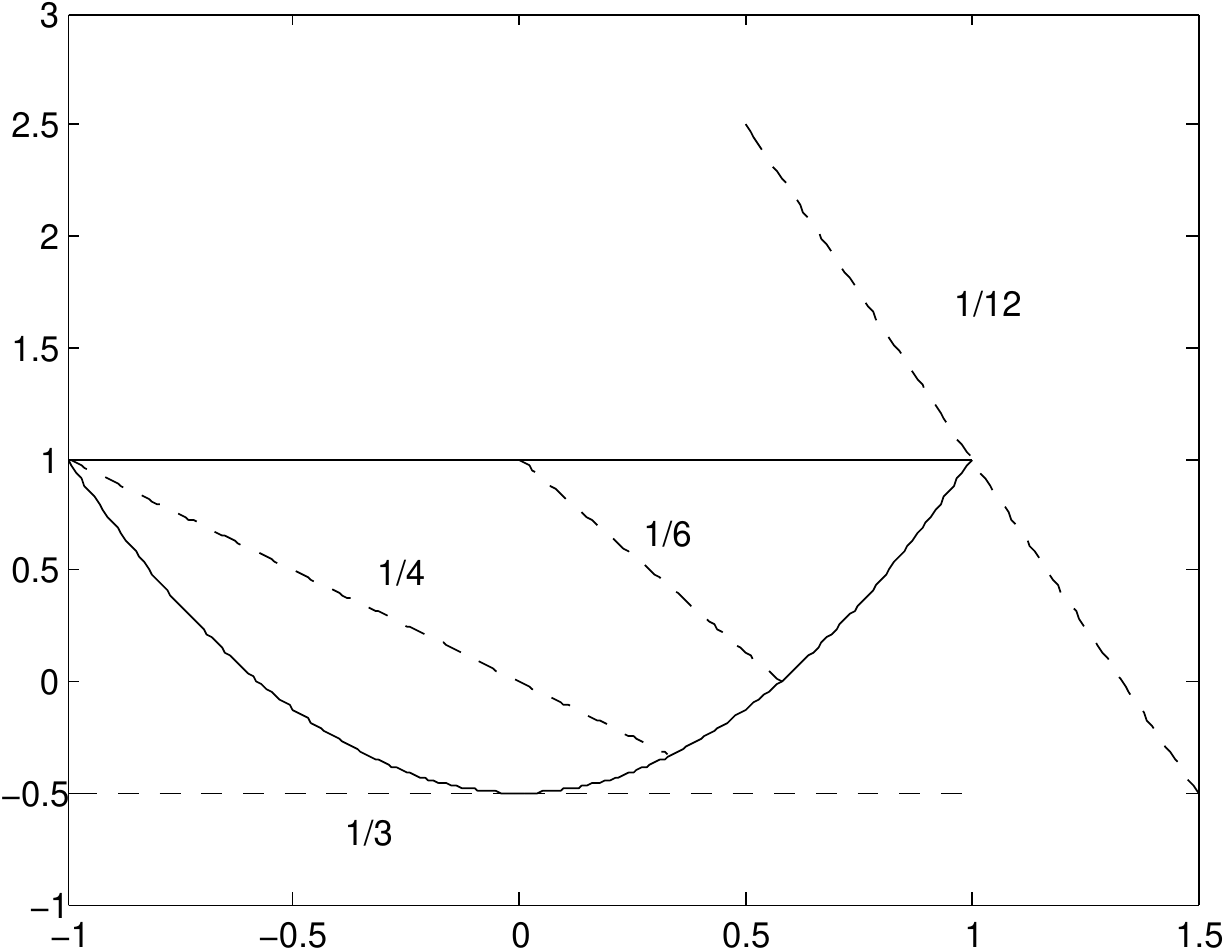}}  
\caption{The solution region for the depolarizing channel satisfying \eqref{posdep}.
  For  $a^2 = \tfrac{1}{12} $ the line \eqref{dep1} is $3x+y = 1$ which yields the
  unique solution $x = y = 1$.    At the EB boundary $a^2 =  \tfrac{1}{6}$
   the  solutions lie on the line from  $(0 ,1)$  to   $(\tfrac{1}{\sqrt{3}},0)$;  and
    at  $a^2 = \tfrac{1}{4} $  on the line from  $(-1,1)$  to   $(\tfrac{1}{3}, -\tfrac{1}{3})$.
   At  $a^2 = \tfrac{1}{3} $ one has only the unique solution  $(0,-0.5)$.} 
    \label{fig:dep} \end{figure}

For the depolarizing channel with $a_k = a$ for $k \neq 0$ and $a_0 =
\sqrt{1 - 3 a^2}$
and the assumption of symmetric solutions $x_k = x$, $y_k = y$,
\eqref{cond}  becomes
\be  \label{dep1}
   2 a  \sqrt{1 - 3 a^2} x + 2 a^2 y   = 1 - 4 a^2.
\ee
When $a^2 = \frac{1}{12} $, \eqref{dep1} becomes
$    \half x + \frac{1}{6} y $ whose only solution in the unit square
is $x = y = 1$, for which \eqref{antimat} is a multiple of a rank
one projection and hence, on the boundary of the cone of
positive semi-definite matrices.   In the entanglement-breaking region
$\tfrac{1}{6}  \leq  a^2  \leq  \tfrac{1}{3} $,
$x = 0, y = \frac{1 - 4a^2}{2 a^2}$ always gives a solution for which
\eqref{antimat} is positive semi-definite.   For the general case,
observe that
when $x_1 = x_2 =x$ and $y_1 = y_2 = y$, \eqref{pos_sig} holds if and
only if  $   1 + y  \geq  x^2+ y^2 $ and $ ( 1 + y - x^2 - y^2)^2 \geq
(x^2 - y^2)^2 + x^2(1-y)^2$.
The latter  inequality is stronger in the unit square, and can be
rewritten as
\be  \label{posdep}
(1+y)^2 - 2(x^2+y^2) -2y^3 + 3 x^2 y^2 - x^2 \geq 0.
\ee
Then for  $\tfrac{1}{12} < a^2  <  \tfrac{1}{3} $ one has a family of
{\bf non-unique}
solutions  corresponding to  the line segment
which satisfies  \eqref{dep1} and lies within the region  in the $xy$-plane
bounded above by the
line $y = 1$ and below by curve for which equality holds
in \eqref{posdep}, as shown in Figure~\ref{fig:dep}.    Thus, we have
recovered the
well-known result  \cite{BDEFMS97,Cerf} that depolarizing channels with
$|\lambda_k|  \leq \frac{2}{3} $ are anti-degradable.   Moreover, we have
shown that, except for $\lambda_k = \frac{2}{3}$ and $\lambda_k = -
\frac{1}{3}$,
the degrading map for $\Phi^C$ is {\em not} unique.

\subsection{Proof of Theorems~\ref{thm:CNG}  and \ref{thm:CNGlambda}} \label{app:qantipf}

To study the general case of unital qubit channels, first consider
the situation in which all $\lambda_k \geq 0$ and all $a_j \geq 0 $.
We will then show that the latter does not involve any loss of 
generality and that channels with some $\lambda_k \leq 0$
situations are either entanglement breaking or can be rotated
into the positive case by conjugating with a $\sigma_j$.

First, observe that  all $\lambda_k  > 0$ implies 
\be
      0 <   \lambda_i + \lambda_j = 2 (a_0^2 - a_k^2)  
\ee
with $\{ i,j,k \}$  any permutation of $\{ 1,2,3 \}$.    Therefore, $a_0^2  > a_k^2$
for $k = 1,2,3$.    Combing this with our assumption that all $a_k \geq 0$,
we can conclude that $a_0 \geq a_k$ for $k = 1,2,3$.

Next observe that the requirement that $x_k, y_k$ lie in the unit square,
implies that the absolute value of the LHS of \eqref{cond} is 
bounded above by $2 a_0 a_k + 2 a_i a_j$.
Thus, a necessary condition for anti-degradability is that
\be   \label{ant1}
      \lambda_k = a_0^2 + a_k^2 - a_i^2 - a_j^2 \leq 2 a_0 a_k + 2 a_i  a_j 
\ee
which is equivalent to  $(a_0 - a_k)^2 \leq (a_i + a_j)^2 $.
With the assummption that all $a_j \geq 0$, this implies     
 \be
\label{ant3}
        a_0 \leq a_i + a_j + a_k .
\ee
Substituting \eqref{ant3} into \eqref{ant1}  and using
$a_0^2 = 1 - a_i^2 - a_j^2 - a_k^2$ gives
\eqref{CNG} as a necessary condition for antidegradability in the case 
$\{ i,j,k \} = \{ 1,2,3 \}$.
In the multiplier picture this becomes  (still assuming all $a_j \geq
0$) \be   \label{CNG:lamb}
    \sum_{k = 1}^3  \Big( 1 - \lambda_k + \sqrt{(1 - \lambda_k)^2 -
(\lambda_i - \lambda_j)^2}        \,  \Big) \geq 2 \ee with all $i,j,k$
distinct.   

To show that  \eqref{CNG:lamb} is  sufficient for anti-degradability,
it is enough to verify that $x_k = y_k =  \frac{ a_0^2 + a_k^2 - a_i^2
- a_j^2}{2 a_0 a_k + 2 a_i a_j}$
yields a CPT degrading map for a Pauli channel with multipliers 
$\lambda_k = a_0^2+a_k^2-a_i^2-a_j^2$.    When all $\lambda_k \geq  0 $,
 and $ a_k \geq 0$, the condition $  0 \leq  x_k  =  y_k   \leq 1$ is equivalent to
  $    (a_0 - a_k)^2 \leq (a_i + a_j)^2 $ which is equivalent to  \eqref{ant3}.
Since  \eqref{CNG:lamb} is equivalent to  \eqref{ant3} when all $a_k > 0$, 
we have shown that it is also sufficient for degradability.

\begin{figure}[h] 
\centerline{\includegraphics[height=8cm]{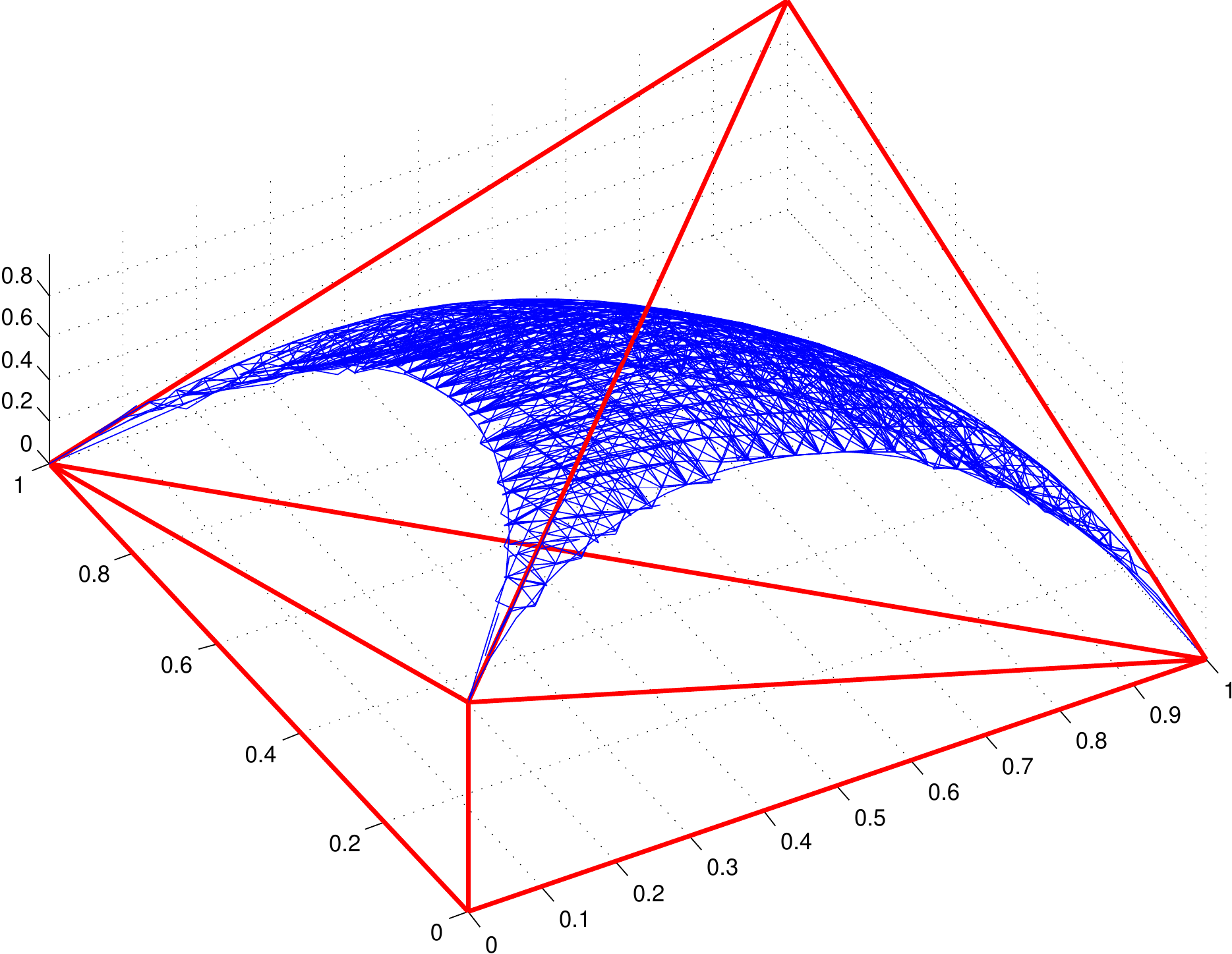}}
\caption{ The hexahedron of unital qubit maps in the sector with all
       $  \lambda_k  \geq 0 $, also showing the boundary of the
       anti-degradable region.   The tetrahedron on the bottom
       corresponds to the subset of EB channels.} \label{fig3}
\end{figure}

Now  a unital qubit channel is independent of the choice of phase
for  the Kraus operators $a_k \sigma_k$.  Hence, its degradability can 
not depend on this phase either, although allowing non-poisitive $a_k$
might yield additional degrading maps.     
  Thus, \eqref{CNG:lamb} is necessary and sufficient  for degradability
when all $\lambda_k > 0$.    The corresponding surface in this quadrant
is shown in Figure~\ref{fig3}.

To complete the proof of Theorem~\ref{thm:CNG} it suffices to observe
that  conjugating with $\sigma_k$   replaces $1,2,3$   in \eqref{CNG}
by $0,i,j$ with
$i,j,k$ distinct in $\{ 1,2,3 \}$.   The corresponding version of
\eqref{CNG:lamb},
has signs modified so that  $\lambda_j \mapsto - \lambda_j$ for $j \neq k$,
and \eqref{CNG:lamb} becomes \eqref{CNG:abs}
\bee 
   \sum_{k = 1}^3  \Big( 1 -| \lambda_k |+ \sqrt{(1 - |\lambda_k|)^2 -
 (|\lambda_i| - |\lambda_j|)^2}        \,  \Big) \geq 2 .\eee
Note that the CP condition \eqref{CP} with $t_j = 0$ implies that the
quantities
under the square root in   \eqref{CNG:lamb} are non-negative.   This remains
true in \eqref{CNG:abs} because changing the sign of two $\lambda_j$
either leaves   $(1 - \lambda_k)^2 -  (\lambda_i - \lambda_j)^2 $ unchanged
or changes it to  $(1 + \lambda_k)^2 -  (\lambda_i + \lambda_j)^2 $,   which  
is also non-negative by  \eqref{CP}.      One way of charactering  the unital
EB class \cite{RuskEB} is that
\eqref{CP} is replaced by the stronger conditions
\be
     (1 \pm \lambda_k)^2 -  (\lambda_i  \mp \lambda_j)^2  \geq 0,
\ee
which is equivalent to $\sum_k |\lambda_k| \leq 1$ and immediately
implies \eqref{CNG:abs}.

Another way of viewing this situation is to observe that interior
of the well-known
tetrahedron of unital qubit maps can be written  as the union of 8
regions:
\begin{itemize}

\item  4 hexahedrons with an even number of $\lambda_n$ negative.
 One corresponds to all  $\lambda_n > 0$; the others can be
 obtained from this by conjugating with $\sigma_k, ~ k =1,2,3$
for which $\lambda_k >  0$ and the remaining two $\lambda_j < 0$.

\item 4 tetrahedrons with an odd number of $\lambda_n$ negative.
 One corresponds to all  $\lambda_n < 0$; the others can be
 obtained from this by conjugating with $\sigma_k, ~ k =1,2,3$
for which $\lambda_k <  0$ and the remaining two $\lambda_j > 0$.
 
\end{itemize}

It was shown in  \cite{RuskEB}  that any channel which remains CP
when $ \lambda_k \mapsto - \lambda_k$ (which is equivalent
to applying the partial transpose to the Choi matrix and conjugating with
a Puali matrix)  is EB.    It follows that  all unital qubit channels with an odd 
number of $\lambda_n$ negative, or any $\lambda_n = 0$, is EB.
Moreover, a unital qubit channel is EB if and only if $\sum_k |\lambda_k| \leq 1$
which implies that \eqref{CNG:abs}.     Thus we have proven 
Theorem~\ref{thm:CNGlambda}.
In the case of channels with an odd number of negative  $\lambda_k$
it can happen that a linear qubit map of the form \eqref{qumap}
    satisfies  \eqref{CNG:abs} without being CP.
 Therefore it is important that the CP condition 
 $ (1 \pm \lambda_k)^2  \geq  (\lambda_i  \pm \lambda_j)^2$ is included 
 in the hypothesis.
 
 Indeed, the astute reader will note that the proof  found it sufficient to 
 consider degrading maps $\Psi$ with $x_k = y_k$.    However, the
 constraints on the degrading map for depolarizing channels in 
 Section~\ref{app:antdep} imply that for $a^2 \approx \tfrac{1}{3}$ no solution
 with $x = \pm y$ exists.   There is no contradiction because for $a^2 > \tfrac{1}{4}$,
 the multiplier $\lambda < 0$ and the assumption that $a_0^2$ is largest no
 longer holds.    This does, however, demonstrate the need to consider the
 four small tetrahedrons with an odd number of $\lambda_n$ negative separately.

\end{document}